\documentclass[12pt]{article}
\pdfoutput=1

\usepackage[
top = 2.5cm, 
bottom = 2.5cm, 
left = 2.5cm, 
right = 2.5cm]{geometry}

\usepackage{jheppub}
\usepackage[sort&compress]{natbib}
\usepackage[utf8]{inputenc}
\usepackage{ifpdf}
\usepackage{braket}
\usepackage{graphicx}
\usepackage{caption}
\usepackage{subcaption}
\usepackage{slashed}
\usepackage{feynmf}
\usepackage{epstopdf}
\usepackage{amsmath, amssymb,url,bm,textgreek,upgreek}
\usepackage[svgnames]{xcolor}
\usepackage{hyperref}
\usepackage{dsfont}
\usepackage{physics}
\usepackage{mathtools}
\hypersetup{linkcolor=DarkBlue,citecolor=DarkBlue, urlcolor=DarkBlue}

\pdfoutput=1
\makeatletter
\def\@fpheader{\vspace{1mm}}
\makeatother

\def\be{\begin{eqnarray}}
\def\ee{\end{eqnarray}}

\def\OO{{\mathcal O}}

\def \bea {\begin{eqnarray}}
\def \eea {\end{eqnarray}}
\def \nn {\nonumber}

\def \k {\kappa}

\def \del {\partial}

\def\EE{\mathcal{E}}

\title{Freedom near Lightcone and  ANEC Saturation}

\author{Kuo-Wei Huang,$^{a,b}$}
\author{Robin Karlsson,$^{c,b}$}
\author{Andrei Parnachev$^{b,d}$}
\author{and Samuel Valach$^{b}$}

\affiliation{$^a$Department of Physics, Boston University, Boston, MA 02215, U.S.A.}
\affiliation{$^b$School of Mathematics and Hamilton Mathematics Institute, 
\setlength{\parskip}{0pt}

~Trinity College, Dublin 2, Ireland}
\affiliation{$^c$CERN, Theoretical Physics Department, CH-1211 Geneva 23, Switzerland}
\affiliation{$^d$Kadanoff Center for Theoretical Physics and Enrico Fermi Institute, 
\setlength{\parskip}{0pt}

~University of Chicago, Chicago IL 60637, U.S.A.}

\preprint{CERN-TH-2022-173}

\abstract{Averaged Null Energy Conditions (ANECs) hold in unitary quantum field theories. 
In conformal field theories, ANECs in states created by the application of the stress tensor to the vacuum 
lead to three constraints on the stress-tensor three-point couplings, depending on the choice
of polarization.
The same constraints follow from considering two-point functions of the stress tensor in a thermal state and 
focusing on the contribution of the stress tensor in the operator product expansion (OPE).
One can observe this in holographic Gauss-Bonnet gravity, where ANEC saturation coincides with the appearance of superluminal signal propagation in thermal states.
We show that, when this happens, the corresponding generalizations of ANECs for higher-spin multi-stress tensor operators with minimal twist are saturated as well and all contributions from such operators to the thermal two-point functions vanish in the lightcone limit. 
This leads to a special near-lightcone behavior of the thermal stress-tensor correlators --
they take the vacuum form, independent of temperature.
}

\begin{document}
\maketitle

\addtolength{\parskip}{0.8 ex}

\newpage
\section{Introduction}

Exploring universal constraints and their consequences in quantum field theories (QFTs) is of great importance. 
The  present paper considers questions related to Averaged Null Energy Conditions (ANECs) which  generally hold in unitary QFTs \cite{Faulkner:2016mzt,Hartman:2016lgu}.   More precisely, we focus on conformal field theories (CFTs) where important examples of ANECs are conformal collider bounds  \cite{Hofman:2008ar}. In this work, we shall pay special attention to the situation where ANECs are saturated, and discuss the connection to stress-tensor correlators at finite temperature.

In the setup of \cite{Hofman:2008ar},  localized states are created by the stress tensor
with three independent polarizations. The energy flux is determined by the three numbers specifying the stress-tensor three-point functions and the positivity of  the energy flux 
results in three constraints on the combinations of these couplings.
Recent advances in CFT techniques (see, $e.g.$, \cite{Rychkov:2016iqz,Simmons-Duffin:2016gjk,Poland:2018epd} for reviews) 
 allowed proving conformal  collider bounds in unitarity CFTs \cite{Hofman:2016awc} (see also \cite{Li:2015itl,Komargodski:2016gci}). 
The bootstrap proof focuses on the lightcone limit of a four-point function with two scalars and two stress-tensor insertions, 
 which is dominated by the stress-tensor exchange. 
The same techniques allow making statements about interference effects in conformal collider bounds and higher-spin ANECs 
\cite{Cordova:2017zej,Meltzer:2017rtf,Meltzer:2018tnm}.
(See \cite{Balakrishnan:2017bjg,Cordova:2017dhq,Kravchuk:2018htv,Delacretaz:2018cfk,Cordova:2018ygx,Ceyhan:2018zfg,Belin:2019mnx,Kologlu:2019mfz,Manenti:2019kbl,Belin:2020lsr,Besken:2020snx,Korchemsky:2021okt,Korchemsky:2021htm,Caron-Huot:2022eqs}
for some examples of recent work devoted to the study of ANECs.)

 In  \cite{Kulaxizi:2010jt}, it was pointed out that one can observe conformal collider bounds by studying two-point functions of the stress tensor 
 (the $TT$ correlators) at finite temperature, using the operator product expansion (OPE) and focusing on the contribution of the stress tensor. 
 Symmetries  imply that the stress-tensor two-point functions at finite temperature have three independent polarizations.
 As explained in \cite{Kulaxizi:2010jt} (see also \cite{Karlsson:2022osn}) the coefficients of the 
  stress-tensor contributions in the lightcone limit for these polarizations are
 precisely proportional to the corresponding ANECs.
 When one of these coefficients vanish, the corresponding ANEC gets saturated.
Here, we ask the following question: can this result be generalized to include the contributions from multi-stress tensor exchanges?

In this paper, via holography  \cite{Maldacena:1997re, Gubser:1998bc, Witten:1998qj}, we adopt Gauss-Bonnet gravity to study ANEC saturations using thermal $TT$ correlators. Gauss-Bonnet gravity and more generally Lovelock theories are useful theoretical laboratories for studying higher-derivative corrections because their equations of motion are of second order.  Our working hypothesis is that Gauss-Bonnet gravity, despite being a special theory, might allow us to identify some universal features of holographic CFTs regardless of what higher-derivative terms are included. Indeed, ANECs manifest themselves via the superluminal propagation of signals in Gauss-Bonnet gravity \cite{Brigante:2007nu,Brigante:2008gz,deBoer:2009pn,Camanho:2009vw,Buchel:2009sk}. (For more recent developments in the holographic aspects of Gauss-Bonnet gravity, see, $e.g.$, \cite{Buchel:2009tt, 
Buchel:2010wf,Cai:2010cv,Bu:2015bwa,Grozdanov:2016vgg, Andrade:2016yzc, Grozdanov:2016zjj,Andrade:2016rln,Grozdanov:2016fkt, Chen:2018nbh,An:2018dbz,Grozdanov:2021gzh}.) 
 While the holographic Gauss-Bonnet theory is not unitary \cite{Camanho:2014apa}, the breakdown of unitarity 
 for small values of the Gauss-Bonnet coupling happens in the small impact
 parameter regime, as opposed to the large impact parameter (lightcone) limit relevant for ANECs.\footnote{This
 can be seen by analyzing corresponding CFT four-point functions in the impact parameter space. See, $e.g.$, \cite{Camanho:2014apa,Kulaxizi:2017ixa,Li:2017lmh,Costa:2017twz,Bonifacio:2017nnt,Afkhami-Jeddi:2018own,Kologlu:2019bco,Caron-Huot:2021enk}. At finite values of the Gauss-Bonnet coupling,  light higher-spin operators  are needed to restore unitarity. Since we do not have control over the full tower of such higher-spin operators, we do not include them in our analysis.}
 This  is why holographic Gauss-Bonnet gravity allows one to observe conformal collider bounds which have a much larger
 degree of universality and apply to all unitary CFTs\footnote{Note that to study the 
 regime of ANEC saturation we need to consider large higher derivative terms in the gravitational lagrangian.
For generic  such terms this would lead to equations of motions which will
be higher than second order and will result in a variety of complications. 
Gauss-Bonnet gravity is special in this regard.
 }.

The results of \cite{Brigante:2007nu,Brigante:2008gz} on superluminal propagation in
 Gauss-Bonnet gravity can be directly connected to the OPE analysis of   \cite{Kulaxizi:2010jt}.
 Consider the integrated $TT$ correlators on $S^1_\beta \times \mathbb{R}^3$: 
\begin{equation}
\label{intcorr}
	G_{\mu\nu,\rho\sigma}(t, z; \beta)=\int_{\mathbb{R}^2}  \dd x\dd y ~ \langle T_{\mu\nu}(t,x,y,z)T_{\rho\sigma}(0)\rangle_\beta,
\end{equation} 
where $\beta=T^{-1}$ is the inverse temperature.
Choosing a particular polarization and expanding the holographic correlator in powers of temperature
one should be able to see that when the corresponding ANEC is saturated, the leading near-lightcone 
$\OO(\beta^{-4})$ term in the expansion vanishes.
We perform the finite-temperature expansion of  (\ref{intcorr}) using the techniques developed in  \cite{Karlsson:2022osn} and
confirm this expectation.
We then consider the subsequent $\OO(\beta^{-8})$ terms in the expansion and extract 
 the contribution of the spin-4 double-stress tensor operator.
We observe that when  a spin-2 ANEC is saturated, for the same choice of polarization the spin-4 ANEC is also saturated
 and the leading near-lightcone $\OO(\beta^{-8})$ term in the expansion vanishes as well.

 Does this pattern persist to all orders in the temperature expansion?
 Since multi-stress tensor operators of highest spin (for a given conformal dimension) govern the near-lightcone behavior, to answer this question we need to study the near-lightcone regime.  
We analyze the near-lightcone thermal $TT$ correlators to all orders\footnote{We do this by generalizing the approach of \cite{Fitzpatrick:2019zqz}, where near-lightcone scalar correlators were studied,  to the stress-tensor case.}
and observe that once
a spin-2 ANEC for a certain polarization is saturated,  the leading-lightcone limit of the correlator for this polarization 
takes the vacuum form and is completely independent of the temperature.
Hence, all spin-$2 k$ ANECs for multi-stress tensor operators $[T_{\mu\nu}]^k$ of maximal spin are saturated.

It has been observed that free theories saturate conformal collider bounds \cite{Hofman:2008ar}.
However it is less obvious whether theories which saturate conformal collider bounds are necessarily free, although
some evidence in this direction was presented in \cite{Zhiboedov:2013opa,Meltzer:2018tnm}.
In this paper we propose a scenario where the theory is ``free" in a limited sense: correlators of the stress-tensor
take a vacuum form for one particular polarization.
We call this behavior ``freedom near lightcone" and observe it in holographic Gauss-Bonnet gravity.

To make contact with the literature, we  read off the double-stress tensor CFT data to subleading order in the $C_T^{-1}$ expansion by comparing the bulk computations to the OPE  in the dual CFT. The leading order mean field theory (MFT) result needs to satisfy consistency conditions. 
These are due to interference effects of the ANEC in states that are superpositions of the stress tensor and double-stress tensors of spin $0,2,4$. For the spin-$0$ double-stress tensor this was shown to impose no constraint on the OPE coefficient \cite{Cordova:2017zej}, while for spin-$2$ and spin-$4$ double-stress tensors interference effects  impose non-trivial constraints on the OPE coefficients \cite{Meltzer:2017rtf,Meltzer:2018tnm}. 
We verify that the MFT coefficients in holographic CFTs are consistent with such interference effects. 
In addition, following \cite{Meltzer:2018tnm},  from the CFT point of view we verify that, using the data obtained from holographic Gauss-Bonnet gravity, the spin-$4$ ANEC is also saturated when the corresponding spin-$2$ ANEC is saturated.

\subsubsection*{Outline}

In the next section, we 
write down the  equations of motion in Gauss-Bonnet (GB) gravity and analyze them using a near-boundary expansion.
Our calculations are done for the four-dimensional CFT case, but we expect to find similar results in other  dimensions. 
In Section 3, we show that, when an ANEC is saturated all higher-spin ANECs for the same polarization are saturated as well and the corresponding $TT$ correlator near the lightcone is reduced to the vacuum form. 
We read off CFT data for the double-stress tensors in the context of GB gravity in Section 4 by performing the conformal block decomposition.  
 Section 5 is devoted to a discussion of conformal collider bounds for states which are linear combinations of stress tensors and double-stress tensors, as well as the study of the spin-$4$ ANEC. We conclude  in Section 6 with a list of future directions.  

\section{Thermal \textit{TT} and Gauss-Bonnet Gravity}\label{Sec:Bulk}

In this section, 
after a brief review of Gauss-Bonnet gravity  we study perturbations of the planar black hole, setting up our notations and introducing the ansatz used to compute the bulk-to-boundary propagators.  
We then discuss the  thermal stress tensor two-point functions for different polarizations and analyze the contributions of the identity, the stress tensor and the double-stress tensors. 
The near-lightcone behavior of the stress-tensor correlators, including an all order analysis, will be discussed in the next section.  

\subsection{A Brief Review on Gauss-Bonnet Gravity}

In the Euclidean signature, we write the five-dimensional Gauss-Bonnet action with a negative cosmological constant as
\begin{align}\label{actionGBshort}
S_{GB}& =\frac{1}{16\pi G}\int\dd^5x\sqrt{g}\left[\frac{12}{L^2}+R+\lambda_{GB}\frac{L^2}{2}\big( R^2_{\mu\nu\lambda\rho}-4R^2_{\mu\nu}+R^2  \big)\right]
\end{align}
where $G$ is the gravitational constant and $\lambda_{GB}$ is the (dimensionless) Gauss-Bonnet coupling. 
Despite having higher curvature terms, the equations of motion resulting from \eqref{actionGBshort} remain second-order PDEs. 
For technical reasons we focus on the planar (large radius) AdS black hole solution:
    \begin{equation}\label{gbbh}
        \dd s^2=\frac{r^2}{L^2}\left(\frac{f(r)}{f_\infty}\dd t^2+\dd \vec{x}^2\right)+\frac{L^2}{r^2}\frac{\dd r^2}{f(r)} \ ,
    \end{equation}
where $f(r)$ and $f_\infty$ are \cite{Boulware:1985wk, Cai:2001dz}
    \begin{align}
        f(r)&=\frac{1}{2\lambda_{GB}}\left[1-\sqrt{1-4\lambda_{GB}\left(1-\frac{\tilde{\mu}}{r^4}\right)}\right]\label{defefss}\ , \\
        f_\infty&=\lim_{r\rightarrow\infty}f(r)=\frac{1-\sqrt{1-4\lambda_{GB}}}{2\lambda_{GB}}\label{finfdef}  \ . 
    \end{align} 
This solution corresponds to a nonsingular black hole in a ghost-free vacuum.  No AdS vacuum exists if $\lambda_{GB}>1/4$. 
 The normalization of the metric is chosen such that the speed of light is one in the dual CFT. 
The parameter $\tilde{\mu}$ and the Hawking temperature $T$ are related in the following way \cite{Cai:2001dz}:
\begin{align}
T={r_+ \over \pi L^2 \sqrt{f_\infty}} \ , ~~~~~ r_{+}^4= \tilde{\mu}
   \end{align}  
where $r_+$ denotes the location of the black-hole horizon.

Taking $\tilde\mu\rightarrow0$ in \eqref{gbbh}, one recovers the AdS vacuum in the Poincar\'{e} coordinates:
    \begin{equation}
        \dd s^2=\frac{r^2}{L^2}\delta_{ab}\dd x^a\dd x^b+\frac{\tilde{L}^2}{r^2}\dd r^2 \ , ~~~~~ \tilde L= {L\over \sqrt{f_\infty} } 
\end{equation}
where $a,b\in\{t,\,x,\,y,\,z\}$ and $\tilde L$ is the AdS curvature scale.  The metric acquires a simpler form
    \begin{equation}\label{vacGBconven}
        \dd s^2=\tilde L\left(\tilde{r}^2\delta_{ab}\dd x^a\dd x^b+\frac{1}{\tilde{r}^2}\dd \tilde{r}^2\right) \ , ~~~ \tilde r={r\over L\tilde L}  
    \end{equation}    using the rescaled coordinate $\tilde r$. 

The central charge $C_T$ of the CFT dual to Gauss-Bonnet gravity is \cite{Buchel:2009sk}
    \begin{equation}\label{CTdefGB}
    C_T=\frac{5L^3}{\pi^3Gf_\infty^{3/2}}(1-2f_\infty\lambda_{GB}) \ .
    \end{equation}
One can relate the parameter $\tilde{\mu}$ to the conformal dimension $\Delta_H$
  of the heavy operator that creates a heavy state  \cite{Karlsson:2020ghx}:
    \begin{equation}\label{eq:defMuTilde}
        \tilde{\mu}=\frac{20}{3 \pi ^4}\left(1-4 \lambda_{GB} +\sqrt{1-4 \lambda_{GB} }\right) \frac{\Delta_H}{C_T} f_\infty^4 \tilde L^4\  .
    \end{equation} 
In the following we will often set $\tilde L=1$, in which case $L= \sqrt{f_\infty}$.

    \subsection{Black Hole Perturbations and Ansatz}

We shall consider a small perturbation $h_{\mu\nu}$ of the black-hole metric \eqref{gbbh} and restrict ourselves to the case where $h_{\mu\nu}$ does not depend on the coordinates $x$ and $y$.
According to the representations under the rotations in the $xy$-plane, the fluctuations can be classified into three channels:
\begin{align}
\text{Scalar channel  (spin 2)}&:\qquad h_{\alpha\beta}-\delta_{\alpha\beta}(h_{xx}+h_{yy})/2 \\
\text{Shear channel (spin 1)}&:\qquad h_{tx},\,h_{ty},\,h_{zx},\,h_{zy},\,h_{rx},\,h_{ry} \\
\text{Sound channel (spin 0)}&:\qquad h_{tt},\,h_{tz},\,h_{zz},\,h_{rr},\,h_{tr},\,h_{zr},\,h_{xx}+h_{yy} \ . 
\end{align}
The linearized equations of motion then can be studied separately for each spin, as different representations do not mix.
For each channel, we adopt a quantity $Z$ invariant under diffeomorphisms \cite{Buchel:2009sk}:
    \begin{align}
    &Z_{\rm{scalar}}=H_{xy},\label{Z3Def}\\
    &Z_{\rm{shear}}=\partial_zH_{tx}-\partial_tH_{xz}\label{Z1Def}\ , \\
    &Z_{\rm{sound}}=\frac{2f}{f_\infty}\partial_z^2H_{tt}-4\partial_t\partial_zH_{tz}+2\partial_t^2H_{zz}\nn\\
&~~~~~~~~~~~ -\left(\left(\frac{f}{f_\infty}+\frac{r\partial_rf}{2f_\infty}\right)\partial_z^2+\partial_t^2\right)\left(H_{xx}+H_{yy}\right),\label{Z2Def} 
    \end{align}
where  
    \begin{equation}
    H_{tt}=\frac{L^2}{r^2}\frac{f_\infty}{f(r)}h_{tt}\ , \quad H_{ti}=\frac{L^2}{r^2}h_{ti}\ , \quad H_{ij}=\frac{L^2}{r^2}h_{ij}\ ,  \quad i,\,j\in\{x,y,z\}     \ . 
    \end{equation}
The equations of motion for all three channels have the following form \cite{Buchel:2009sk}: 
    \begin{equation}\label{eomsschm}
        \partial_{\tilde{r}}^2Z+C^{(1)}\partial_{\tilde{r}}Z+C^{(0)}Z =0 \ ,
    \end{equation}
where $C^{(1)}$ and $C^{(0)}$ are differential operators. 
In the scalar channel, they are given by 
    \begin{align}
        C_{\rm{scalar}}^{(1)}&=\frac{4}{f^2 (\kappa +1)^2 \tilde{r}^4}\partial_t^2+\frac{6 f \left(\kappa ^2-1\right) \left(f \left(\kappa ^2-1\right)+4\right)-16 \kappa ^2+24}{f (\kappa +1) \tilde{r}^4 \left(f \left(\kappa ^2-1\right)+2\right)^2}\partial_z^2\ , \\
        C_{\rm{scalar}}^{(0)}&=\frac{f \left(f \left(\kappa ^2-1\right) \left(5 f \left(\kappa ^2-1\right)+16\right)+4\right)+16}{f \tilde{r} \left(f \left(\kappa ^2-1\right)+2\right)^2} \ .
    \end{align}where we introduce 
    \begin{equation}\label{defLambda}
     \kappa= \sqrt {1- 4\lambda_{GB}} 
    \end{equation} 
which will help simplify various expressions.
The shear channel has
{
\allowdisplaybreaks
    \begin{align}
    C_{\rm shear}^{(1)}=&\frac{\left(f
   \left(\kappa ^2-1\right)+2\right)^2 \left(f \left(f \left(\kappa ^2-1\right) \left(5 f \left(\kappa ^2-1\right)+16\right)+4\right)+16\right)}{f \tilde{r} \left(f \left(\kappa
   ^2-1\right)+2\right)^2 \left({\partial_t}^2 \left(f \left(\kappa ^2-1\right)+2\right)^2+2 f (\kappa +1) \kappa ^2 {\partial_z}^2\right)}{\partial_t}^2\nonumber\\
   &+\frac{2 f^2 (\kappa +1) \kappa ^2 \left(3 f \left(\kappa ^2-1\right) \left(f \left(\kappa ^2-1\right)+4\right)+8 \kappa ^2+12\right)}{f \tilde{r} \left(f \left(\kappa^2-1\right)+2\right)^2 \left({\partial_t}^2 \left(f \left(\kappa ^2-1\right)+2\right)^2+2 f (\kappa +1) \kappa ^2 {\partial_z}^2\right)}\partial_z^2\ , \\
   C_{\rm shear}^{(0)}=&\frac{4}{f^2 (\kappa +1)^2 \tilde{r}^4}\partial_t^2+\frac{8 \kappa ^2}{f (\kappa +1) \tilde{r}^4 \left(f \left(\kappa ^2-1\right)+2\right)^2}\partial_z^2 \ .
    \end{align}
}The corresponding $C^{(1)}$ and $C^{(0)}$ for the sound channel can be obtained by Fourier transforming and Wick rotating the corresponding expressions in Appendix D of \cite{Buchel:2009sk}. Due to their length, we will not present them here. 

The above equations of motion are difficult to analyze in general. However, 
using the  techniques developed in \cite{Fitzpatrick:2019zqz, Karlsson:2022osn}, we can solve these equations focusing on the regime 
    \begin{equation}\label{limitoss}
	\tilde{r}\rightarrow\infty\qquad\text{with}\qquad \tilde{r}t,\, \tilde{r}z\,\,\,\text{fixed} \ ,
    \end{equation}
which corresponds to the OPE limit on the boundary. Introducing new variables 
    \begin{align}
        \rho=\tilde r z \ , ~~~~~~~~ w^2=1+\tilde{r}^2t^2+\tilde{r}^2z^2 \ ,
    \end{align} 
 the limit \eqref{limitoss} can be rephrased as $\tilde{r}\rightarrow\infty$ with $\rho$ and $w$ held fixed.
We write the bulk-to-boundary propagators $\mathcal{Z}$ as
\begin{equation}
	Z(t,z,r)=\int\dd t'\dd z'\mathcal{Z}(t-t',z-z',r)\hat{Z}(t',z') 
\end{equation}
where the invariant $\hat{Z}$ is (up to derivatives, as will be explained on separated channels below) the boundary value of $Z$. 
In the near-boundary, OPE expansion, we can solve the equations of motion by taking
\begin{equation}\label{thetheansatz}
	\mathcal{Z}=\mathcal{Z}^{AdS}\left(1+\frac{1}{\tilde{r}^4}\left(G^{4,1}+G^{4,2}\log \tilde{r}\right)+\frac{1}{\tilde{r}^8}\left(G^{8,1}+G^{8,2}\log \tilde{r}\right)+\ldots\right),    
\end{equation}
\vspace{-0.8cm}
\begin{align}
	&G^{4,j}=\sum_{m=0}^2\sum_{n=-2}^{4-m}(a^{4,j}_{n,m}+b^{4,j}_{n,m}\log w)w^n\rho^m \ , \\
	&G^{8,j}=\sum_{m=0}^6\sum_{n=-6}^{8-m}(a^{8,j}_{n,m}+b^{8,j}_{n,m}\log w)w^n\rho^m \ .
\end{align}

One can check\footnote{See also \cite{Liu:1998bu,Buchel:2009sk}.} that the bulk-to-boundary propagators in pure AdS vacuum  $\mathcal{Z}^{AdS}$ for various choices of sources $\hat{H}_{\mu\nu}$ (which are the boundary values of $H_{\mu\nu}$)
are given by
{
\allowdisplaybreaks
\begin{align}
&\hat{H}_{xy}:~~~~~ \mathcal{Z}_{\rm{scalar}}^{AdS} =\frac{2\tilde{r}^2}{\pi w^6} \ , \\
&\hat{H}_{tx}:~~~~~ \mathcal{Z}_{\rm{shear}}^{AdS}=-\frac{12\tilde{r}^3\rho}{\pi w^8} \ , \\
&\hat{H}_{xz}:~~~~~ \mathcal{Z}_{\rm{shear}}^{AdS} =\frac{12\tilde{r}^3}{\pi w^8} \sqrt{w^2-\rho^2-1} \ ,  \\
&\hat{H}_{tz}:~~~~~ \mathcal{Z}_{\rm{sound}}^{AdS}=-\frac{384\tilde{r}^4\rho}{\pi w^{10}} \sqrt{w^2-\rho^2-1} \ , \\
&\hat{H}_{tt}:~~~~~ \mathcal{Z}_{\rm{sound}}^{AdS}=-\frac{24\tilde{r}^4}{\pi w^{10}} (w^2-8\rho^2) \ ,\\
&\hat{H}_{xx}:~~~~~ \mathcal{Z}_{\rm{sound}}^{AdS}=- \frac{24\tilde{r}^4}{\pi w^{10}} (3w^2-4) \ ,\\
&\hat{H}_{zz}:~~~~~ \mathcal{Z}_{\rm{sound}}^{AdS} =\frac{24\tilde{r}^4}{\pi w^{10}} \big(7w^2-8(1+\rho^2)\big) \ ,
\end{align}}where we have expressed these results in terms of  variables $\rho$ and $w$.  
Inserting \eqref{thetheansatz} into the equations of motion, we will obtain $a_{n,m}^{k,j}$ and $b_{n,m}^{k,j}$ for different channels. 

    \subsection{Holographic Thermal $TT$ Correlators}

Let us first recall the holographic dictionary before proceeding to the computation of the stress-tensor correlators.   
The quadratic part of the on-shell action for a general perturbation $H_{\mu\nu}$ in the AdS vacuum was calculated in \cite{Buchel:2009sk}.  
By restricting $H_{\mu\nu}$ to be independent of $x$ and $y$, one has\footnote{Note the sign difference compared to Eq. (3.11) in \cite{Buchel:2009sk}, which is related to the presence of a minus sign in the stress-tensor two-point function defined later in \eqref{varCon}.}  
    \begin{equation}\label{finalactiongbda}
       I=\frac{\pi^2C_T}{320}\int_{\partial M}\dd^4x\,\tilde{r}^5H_{\mu\nu}(t,z,\tilde{r})\partial_{\tilde{r}}H_{\mu\nu}(t,z,\tilde{r}) \ .
    \end{equation}
The action for the perturbations $H_{\mu\nu}$ of the black-hole metric \eqref{gbbh} has the form \eqref{finalactiongbda} plus terms higher-order  in $1/\tilde{r}$ that vanish in the $\tilde{r}\rightarrow\infty$ limit.     
Thus, using \eqref{finalactiongbda} and the definitions \eqref{Z3Def}-\eqref{Z1Def}, one finds the corresponding on-shell actions for invariants to be
    \begin{align}
 I_{\rm scalar}&=\frac{\pi^2C_T}{160}\lim_{\tilde{r}\rightarrow\infty}\int\dd ^4x\tilde{r}^5Z_{\rm scalar}(t,z,\tilde{r})\partial_{\tilde{r}}Z_{\rm scalar}(t,z,\tilde{r}) \ ,  \label{relaction3} \\
        I_{\rm shear}&=\frac{\pi^2C_T}{160}\lim_{\tilde{r}\rightarrow\infty}\int\dd^4x\frac{\tilde{r}^5}{\partial_t^2+\partial^2_z}Z_{\rm shear}(t,z,\tilde{r})\partial_{\tilde{r}}Z_{\rm shear}(t,z,\tilde{r})\ ,\label{relaction1}\\
        I_{\rm sound}&=-\frac{\pi^2C_T}{1920}\lim_{\tilde{r}\rightarrow\infty}\int\dd^4x\frac{\tilde{r}^5}{(\partial_t^2+\partial_z^2)^2}Z_{\rm sound}(t,z,\tilde{r})\partial_{\tilde{r}}Z_{\rm sound}(t,z,\tilde{r})\label{relaction2} \ . 
    \end{align}

\subsubsection{Scalar Channel}

In the simplest case with only the source $\hat{H}_{xy}$ turned on, we have 
    \begin{equation}\label{legoZ3}
        Z_{\rm{scalar}}(t,z,\tilde{r})=\int\dd t'\dd z'\mathcal{Z}_{\rm{scalar}}^{(xy)}(t-t',z-z',\tilde{r})\hat{H}_{xy}(t',z') \ ,
    \end{equation}
where the superscript index of the bulk-to-boundary propagator $\mathcal{Z}_{\rm{scalar}}^{(xy)}$ indicates the non-zero sources.

After inserting \eqref{thetheansatz} into \eqref{eomsschm} for this channel, we determine $a_{n,m}^{k,j}$ and $b_{n,m}^{k,j}$.
 We expand the solution near the boundary:
    \begin{equation}\label{Z3expans}
        \mathcal{Z}_{\rm{scalar}}^{(xy)}(t,z,\tilde{r})=\delta^{(2)}(t,z)+\frac{1}{\tilde{r}^4}\zeta_{\rm{scalar}}^{(xy)}(t,z)+\ldots 
    \end{equation}
where the dots represent contributions analytic in $t$ and $z$ of order $\mathcal{O}(\tilde{r}^{-6})$ and 
subleading contact terms $\sim \mathcal{O}(\tilde{r}^{-2})$ of the schematic form $\partial^n\delta^{(2)}/\tilde{r}^n$. 
Plugging \eqref{legoZ3} and \eqref{Z3expans} into \eqref{relaction3} and taking the limit $\tilde{r}\rightarrow\infty$ gives 
    \begin{equation}
        I_{\rm{scalar}}=-\frac{\pi^2C_T}{40}\int\dd^2x\dd^2x'\zeta_{\rm{scalar}}^{(xy)}(x-x')\hat{H}_{xy}(x)\hat{H}_{xy}(x') \ ,
    \end{equation}
where $x=\{t,z\}$ and $x'=\{t',z'\}$. The CFT correlator can be obtained via 
    \begin{equation}\label{varCon}
        G^{(bulk)}_{xy,xy}=\expval{T_{xy}(t,z)T_{xy}(0,0)}_\beta=-\frac{\delta^2I_{\rm{scalar}}}{\delta\hat{H}_{xy}(t,z)\delta\hat{H}_{xy}(0,0)}=\frac{\pi^2C_T}{20}\zeta^{(xy)}_{\rm{scalar}}(t,z) \ ,
    \end{equation}
where the superscript ``bulk" indicates that these correlators are computed via holography.  Order-by-order in $\tilde{\mu}$, we obtain 
{
\allowdisplaybreaks
\begin{align}
    G_{xy,xy}^{(bulk)}\Big|_{\tilde{\mu}^0}\!=&\frac{\pi  C_T}{10 \left(t^2+z^2\right)^3} \ , \label{Gxy0}\\
	G_{xy,xy}^{(bulk)}\Big|_{\tilde{\mu}^1}\!=&(5 \kappa -4) \frac{\pi  C_T\tilde{\mu }  (t^2-z^2)}{50 \kappa ^2 (\kappa +1) L^8 \left(t^2+z^2\right)^2}\ , \label{mu1res}\\
    G_{xy,xy}^{(bulk)}\Big|_{\tilde{\mu}^2
    }\!=&\frac{\pi  C_T \tilde{\mu }^2}{1050 \kappa ^4 (\kappa +1)^2 L^{16} \left(t^2+z^2\right)} \Big[3 \left(t^2+z^2\right) \big((\kappa  (89 \kappa -206)\nonumber\\
    &+122) t^2+(\kappa  (809 \kappa -1698)+890) z^2\big) \log \left(t^2+z^2\right)\nonumber\\
    &-2 z^2
   \left(15 (\kappa  (89 \kappa -206)+122) t^2+(5 \kappa  (197 \kappa -506)+1606) z^2\right)\Big]\nonumber\\
   &+\frac{1}{10} \pi  C_T \left(a^{8,1(xy)}_{8,0} \left(t^2-7 z^2\right)-6 z^2 a^{8,1(xy)}_{6,0}\right),\label{Gxy2}
    \end{align}
}where, similar to the Einstein gravity case \cite{Karlsson:2022osn}, the coefficients $a^{8,1(xy)}_{8,0}, a^{8,1(xy)}_{6,0}$ remain undetermined.  In the limit $\kappa\rightarrow1$, $i.e.$, $\lambda_{GB}\rightarrow0$, these correlator results reduce to those in the Einstein gravity case, as they must. 

    \subsubsection{Shear Channel}\label{shearHoloss}
    
When the source $\hat{H}_{tx}$ is turned on, we have 
    \begin{equation}\label{legoZ1}
        Z_{\rm{shear}}(t,z,\tilde{r})=\int\dd t'\dd z'\mathcal{Z}_{\rm{shear}}^{(tx)}(t-t',z-z',\tilde{r})\hat{H}_{tx}(t',z') \ , 
    \end{equation}
    \begin{equation}\label{Z1expans}
        \mathcal{Z}_{\rm{shear}}^{(tx)}(t,z,\tilde{r})=\partial_z\delta^{(2)}(t,z)+\frac{1}{\tilde{r}^4}\zeta_{\rm{shear}}^{(tx)}+\ldots \ . 
    \end{equation}
After solving for the corresponding equation of motion, we insert the solution in  \eqref{relaction1} and take the second variational derivative with respect to the source $\hat{H}_{tx}$. We have 
    \begin{equation}
        G_{tx,tx}^{(bulk)}=\frac{\pi^2C_T}{20}\frac{\partial_z}{\partial_t^2+\partial_z^2}\zeta_{\rm{shear}}^{(tx)} \ .
    \end{equation}
The explicit results, order-by-order in $\tilde{\mu}$, are given by 
{
\allowdisplaybreaks
\begin{align}
	G_{tx,tx}^{(bulk)}\Big|_{\tilde{\mu}^0}=&-\frac{1}{\partial_t^2+\partial_z^2}\frac{3 \pi  C_T \left(t^2-7 z^2\right)}{5 \left(t^2+z^2\right)^5}\label{eq:Bulktxtx0}\  , \\
	G_{tx,tx}^{(bulk)}\Big|_{\tilde{\mu}^1}=&-(\kappa -2)  \frac{1}{\partial_t^2+\partial_z^2}\frac{3 \pi  C_T \tilde{\mu } \left(t^4-6 t^2 z^2+z^4\right)}{100 \kappa ^2 (\kappa +1) L^8 \left(t^2+z^2\right)^4}\label{eq:Bulktxtx1}\ , \\
	G_{tx,tx}^{(bulk)}\Big|_{\tilde{\mu}^2}=&-\frac{1}{\partial_t^2+\partial_z^2}\Bigg[\frac{\pi  C_T \tilde{\mu }^2}{2100 \kappa ^4 (\kappa +1)^2 L^{16} \left(t^2+z^2\right)^3}\Big(-6 (\kappa  (105 \kappa \nonumber\\
	&-388)+60) t^4 z^2-24 (\kappa  (33 \kappa -160)+60) t^2 z^4\nonumber\\
	&+3 (\kappa  (97 \kappa -156)+100) \left(t^2+z^2\right)^3 \log \left(t^2+z^2\right)\nonumber\\
	&+2 (\kappa  (55 \kappa
   +212)+4) z^6\Big)+\frac{3}{5} \pi  C_T a^{8,1(tx)}_{8,0}\Bigg]\label{eq:Bulktxtx2} \ . 
\end{align}
}The coefficient $a^{8,1(tx)}_{8,0}$ is not determined by the near-boundary analysis. These results  in the limit $\kappa\rightarrow1$ agree with the Einstein gravity case.

    \subsubsection{Sound Channel}

The sound-channel computation becomes rather cumbersome. 
We focus on the case with the source $\hat{H}_{tz}$ turned on.  
An analogous analysis gives 
    \begin{equation}
        G_{tz,tz}^{(bulk)}=-\frac{\pi^2C_T}{60}\frac{\partial_t\partial_z}{(\partial_t^2+\partial_z^2)^2}\zeta_{\rm{sound}}^{(tz)} \ ,
    \end{equation}
where $\zeta_{\rm{sound}}^{(tz)}$ is the $1/\tilde{r}^4$ term in the near-boundary expansion of the bulk-to-boundary propagator.
Explicit results up to double-stress tensors exchanges are 
{
\allowdisplaybreaks
\begin{align}
	G_{tz,tz}^{(bulk)}\Big|_{\tilde{\mu}^0}=&-\frac{1}{(\partial_t^2+\partial_z^2)^2}\frac{96 \pi  C_T \left(3 t^4-34 t^2 z^2+3 z^4\right)}{5 \left(t^2+z^2\right)^7}\label{tzmu0}\ , \\
	G_{tz,tz}^{(bulk)}\Big|_{\tilde{\mu}^1}=& (3 \kappa -4)  \frac{1}{(\partial_t^2+\partial_z^2)^2}\frac{8 \pi  C_T\tilde{\mu } \left(t^6-15 t^4 z^2+15 t^2 z^4-z^6\right)}{15 \kappa ^2 (\kappa +1) L^8 \left(t^2+z^2\right)^6}\label{tzmu1}\ , \\
	G_{tz,tz}^{(bulk)}\Big|_{\tilde{\mu}^2}=&-\frac{1}{(\partial_t^2+\partial_z^2)^2}\frac{8 \pi  C_T \tilde{\mu }^2 }{1575 \kappa ^4 (\kappa +1)^2 L^{16} \left(t^2+z^2\right)^5}\Big[(9 \kappa  (61 \kappa -134)\nonumber\\
	&+790) t^8-4 (9 \kappa  (283 \kappa -685)+3970) t^6 z^2+10 (3 \kappa  (185 \kappa \nonumber\\
	&-552)+1090) t^4 z^4+4 (15 \kappa  (59 \kappa -139)+1222) t^2
   z^6\nonumber\\
	&+(3 (98-25 \kappa ) \kappa -154) z^8\Big] \ .  \label{tzmu2}
\end{align}
}Again, these results in the $\kappa\rightarrow1$ limit are consistent with the Einstein gravity case.

\section{Near-Lightcone Dynamics}\label{sectionLCCCB}

In this section we take the near-lightcone limit of the expressions discussed in the previous section.
We observe that when the conformal collider bounds are saturated, the near-lightcone behavior of
$\OO(\beta^{-4})$ terms (coming from the  stress-tensor contribution to the $TT$ OPE) and 
$\OO(\beta^{-8})$ terms (coming from the spin-4 double-stress tensor contribution to the $TT$ OPE) vanishes.
We subsequently provide an all-order analysis by taking the lightcone limit in the bulk equations of motion. 

\subsection{Thermal $TT$ Correlators near the Lightcone}  

We define the lightcone limit by going to the Lorenzian signature and defining $(x^+, x^-)=(i t+z, i t-z)$, and 
then we take $x^- \to 0$.

First consider  $\OO(\tilde{\mu})$ contribution. When the conformal collider bounds are saturated, the corresponding critical values of the GB coupling are
    \begin{align}
\label{kappaCCB}
        \kappa^*_{\rm scalar}=\frac45 \ , ~~~~~~~ \kappa^*_{\rm shear}=2\ , ~~~~~~~  \kappa^*_{\rm sound}=\frac43 \ .
    \end{align}
We immediately observe that the expression \eqref{mu1res}
vanishes, while  \eqref{eq:Bulktxtx1} and \eqref{tzmu1} vanish in the  lightcone limit.

Next, we turn to  $\OO(\tilde{\mu}^2)$ term. 
In a small $x^-$ expansion, we find the thermal correlators \eqref{Gxy2}, \eqref{eq:Bulktxtx2}, and \eqref{tzmu2} have the following behaviour:
{\small{
	\allowdisplaybreaks
    \begin{align}
        G_{xy,xy}^{(bulk)}(x^+,x^-)\Big|_{\tilde{\mu}^2} \underset{x^-\to 0}{=}&
-\frac{(5 \kappa -4)^2 \pi  C_T  (x^+)^3 \tilde{\mu }^2}{600 \kappa ^4 (\kappa +1)^2 L^{16} x^-}
-\frac{\pi  C_T (x^+)^2}{2100 \kappa ^4 (\kappa +1)^2 L^{16}} 
\Big( \tilde{\mu }^2  \big(5 \kappa  (197 \kappa -506)\nonumber\\  &-6 (\kappa  (180 \kappa -373)+192) \log (- x^+ x^-)+1606\big)\nonumber\\
        &+105 \kappa ^4 (\kappa +1)^2 L^{16}
\left(3 a^{8,1 (xy)}_{6,0}+4 a^{8,1 (xy)}_{8,0}\right)\Big)
+\order{x^-}\label{bulkLCxy1}\ , \\
        G_{tx,tx}^{(bulk)}(x^+,x^-)\Big|_{\tilde{\mu}^2} \underset{x^-\to 0}{=}&
\frac{1}{\partial_+\partial_-}\bigg(\frac{ (\kappa -2)^2 17 \pi  C_T (x^+)^3 \tilde{\mu }^2}{33600 \kappa ^4 (\kappa +1)^2 L^{16} (x^-)^3}\nonumber\\
        &+\frac{\pi  C_T \big((204-73 \kappa ) \kappa -76\big) (x^+)^2 \tilde{\mu }^2}{11200 \kappa ^4 (\kappa +1)^2
   L^{16} (x^-)^2}+{\cal O}\big({1\over x^-}\big)\bigg)\label{bulkLCtx1}\ , \\
        G_{tz,tz}^{(bulk)}(x^+,x^-)\Big|_{\tilde{\mu}^2} \underset{x^-\to 0}{=}
&\,\,-\frac{1}{\partial_+^2\partial_-^2}\bigg(\frac{ (4-3 \kappa )^2 11 \pi  C_T (x^+)^3 \tilde{\mu }^2}{6300 \kappa ^4 (\kappa +1)^2 L^{16} (x^-)^5}\nonumber\\
        &+\frac{4 \pi  C_T\big(3 \kappa  (52 \kappa -125)+236\big) (x^+)^2 \tilde{\mu }^2}{8400 \kappa ^4 (\kappa
   +1)^2 L^{16} (x^-)^4}+{\cal O}\big({1\over (x^-)^{3}}\big)\bigg) \ .\label{bulkLCtz1}
    \end{align}
}}We see that the leading lightcone contributions all vanish at the corresponding critical values of the GB coupling.  In the expressions above, we keep the subleading lightcone limit terms which remain non-zero.  

\subsection{Reduced Equations of Motion}

To give an all-order proof, we derive {\it the reduced equations of motion}.  This method was developed in the study of the scalar correlator in $d>2$ holographic CFTs \cite{Fitzpatrick:2019zqz}, but the method works also for the stress-tensor correlators. The basic idea is to identify a bulk limit which isolates the largest spin (or lowest-twist) contributions, corresponding to the largest power of $\rho$ in the ansatz \eqref{thetheansatz}, with $w$ fixed.  
More precisely, starting with the equations of motions written in variables $(\tilde r, w, \rho)$, we perform a change of variables
\begin{align}
(\tilde r, w, \rho)~~ \to  ~~(\tilde r, w, v= {\rho\over \tilde r^2})
\end{align} 
and write
\begin{align}
{\cal Z}(\tilde r,w,\rho) ~~ \to  ~~ {\cal Z}^{\rm AdS} \Big(Q(w,v)+ \bar Q(w,v) \log(\tilde r) \Big) \equiv   {\cal Z}^{\rm AdS}  Q_{\text{tot}}
\end{align} 
where, as before, ${\cal Z}^{\rm AdS}$ is the pure AdS solution. 
In the new variables, the lightcone limit corresponds to taking the large $\tilde r$ limit with $v$ fixed. Subleading terms are suppressed at large $\tilde r$.   
Functions $Q$ and $\bar Q$ determine all the information about the near-lightcone stress-tensor correlators.

\vspace{2mm}

{\noindent{\bf{Scalar Channel:}}}  In this simplest case, we obtain the reduced equation of motion in the form
\begin{align}
\big(\k- {4\over 5} \big)  \tilde\mu  \Theta_1+ \Theta_0 = 0 
\end{align} 
where 
{\small\begin{align}
&{\Theta_1\over 10  v^2 } =  \big(w^2 \del^2_w -13 w \del_w   +48 \big) Q_{\text{tot}}\ , \\
&{\Theta_0\over L^8w^2\k^2\left(\k+1\right)}=\Big((1-w^2 )w^2\del^2_w - w^2 v^2 \del^2_v +2 (w^2  -2 ) v w  \del_v \del_w+ (24-5 w^2) v \del_v \nn\\
&~~~~~~~~~~~~~~~~~~~~~ +(3 w^2-5)w\del_w \Big)  Q_{\text{tot}}
+ \Big( 2(1-w^2) w \del_w +2  w^2 v \del_v +4(w^2-3)\Big) \bar Q  \ .
\end{align}}Here, without solving the equation of motion, we observe that the $\tilde\mu$ dependence disappears if $\kappa=\frac45$
and the solution takes the vacuum form.   
This phenomenon does not persist in the subleading lightcone limit. 
Near the lightcone, one may define a parameter which vanishes when the corresponding ANEC is saturated: 
\begin{align}
\mu_{\rm{eff}({\rm scalar})}=  \big(\k- {4\over 5} \big) \tilde\mu \ .
\end{align}  

\vspace{2mm}

{\noindent{\bf{Shear Channel:}}} There are two sources in the  shear channel. 
In both cases, we find that the reduced  equations of motion can be written as
\begin{align}
\mu^2_{\rm{eff}({\rm shear})}\Theta_{2({\rm shear})} + \mu_{\rm{eff}({\rm shear})}\Theta_{1({\rm shear})}+ \Theta_{0({\rm shear})}= 0  \ , ~~~
\mu_{\rm{eff}({\rm shear})}=  \big(\k- 2 \big) \tilde\mu \ .
\end{align}  
For instance, with the source $\hat H_{tx}$ turned on, we obtain 
{\footnotesize{\begin{align}
{ \Theta_{2({\rm shear})}\over 4  v^4  } =& 
\Big[w^2 \left(w^2 \del^4_w-38 w \del^3_w+591 \del^2_w\right)-4431 w \del_w +13440  \Big]Q_{\text{tot}}\ , 
\end{align}}
\vspace{-7mm}
{\footnotesize{\begin{align}
{\Theta_{1({\rm shear})}\over 2  \kappa ^2 (\kappa +1) L^8 v^2}
=&\Big[2 w^3 \Big( \left(w^2-1\right) w^2  \del^3_w  -v w^3  \del_v \del^2_w     +
(27-16 w^2) w \del^2_w +17 v w^2 \del_v \del_w \nn\\
&~~~~~ +  \left(80 w^2-267\right) \del_w-80 v w  \del_v \Big)-160w^2 \left(w^2-12\right)\Big]\bar Q\nn
\end{align}}
\vspace{-7mm}
{\footnotesize{\begin{align}
 &+\Big[\left(w^4-1\right) w^4 \del^4_w    -2  \left(w^2-3\right)v w^5 \del_v \del^3_w +v^2 w^6  \del^2_v \del^2_w  
-2\left(7 w^4+6 w^2-19\right)  w^3  \del^3_w
 \nn\\
& ~~~~~ -17 v^2 w^5  \del^2_v \del_w  
+ 3  \left(11 w^2-54\right) v w^4 \del_v \del^2_w
-\Big(591 - 324 w^2 - 48 w^4 \Big)w^2 \del^2_w \nn\\
&~~~~~+80 w^4   v^2 \del^2_v 
+3  \left(534-59 w^2\right)v w^3 \del_v \del_w  
+3  \left(64 w^4-1068 w^2+1477\right)w \del_w \nn\\
& ~~~~~+240  \left(w^2-24\right)  w^2 v\del_v  -240 \left(5 w^4-48 w^2+56\right)  \Big] Q_{\text{tot}}\ , \\
{\Theta_{0({\rm shear})}\over \kappa ^4 (\kappa +1)^2 L^{16} w^2}
=& \Big[2 (w^2-1)^2 w^3 \del^3_w-4 v^2 w^5 \del_w \del^2_v +2(w^2-1) vw^4 \del_v \del^2_w\nn\\
& -(32 w^6-86 w^4+54 w^2) \del^2_w+32 w^4 v^2  \del^2_v  + 2(17-7 w^2) vw^3 \del_v \del_w \nn\\
& +2 (94 w^4-347 w^2+267) w \del_w-160   w^2 v \del_v -32 (12 w^4-65 w^2+60)\Big]\bar Q\nn
\end{align}
\vspace{-8mm}
\begin{align}
&+ \Big[(w^2-1)^2 w^4 \del^4_w + 2 v^3 w^5 \del_w \del^3_v +(7-3 w^2) v^2 w^4  \del^2_w\del^2_v+4 (w^2-1) v w^3  \del^3_w \del_v \nn\\
&
~~~ -8 \left(2 w^4-5 w^2+3\right) w^3   \del^3_w  -16 v^3 w^4  \del^3_v  + \left(7 w^4-99 w^2+108\right) v w^2 \del^2_w \del_v  + \left(31 w^2-119\right) v^2 w^3\del_w \del^2_v \nn\\
&
~~~ + \left(106 w^4-311 w^2+213\right) w^2 \del^2_w  -80 v^2 w^2 \left(w^2-7\right) \del^2_v -3  \left(27 w^4-305 w^2+356\right)vw \del_w \del_v \nn\\
&
~~~ -(320 w^4-967 w^2+693 ) w \del_w +16 \left(16 w^4-195 w^2+240\right) v \del_v +80 \left(4 w^2-5\right) w^2  \Big] Q_{\text{tot}} \ . 
\end{align}}}}}The $\tilde\mu$ corrections are suppressed in the lightcone limit when $\kappa=2$.  The expressions for another shear-channel source are similar -- see Appendix \ref{Appendixxzxz}.

\vspace{2mm}

{\noindent{\bf{Sound Channel:}}}  The sound-channel reduced  equations of motion are rather complicated and we do not   include them here.  
After a tedious computation, we are able to verify that, when the corresponding ANEC is saturated, $i.e.$, $\k={4\over 3}$, the pure AdS solutions for all sources solve the sound-channel reduced  equations of motion.

\section{Conformal Block Decomposition}\label{CFTsection}  

In this section, we decompose the stress-tensor two-point function using the stress-tensor  OPE. By matching against the bulk results in Section \ref{Sec:Bulk}, we extract the corresponding CFT data of multi-stress tensors, including their OPE coefficients. 
This section follows closely Section 4 and Appendix C in \cite{Karlsson:2022osn}.  
In order to compare against the bulk results, we study the $TT$ correlators on $S^1_\beta \times \mathbb{R}^3$ integrated over the $xy$-plane, $i.e.$, \eqref{intcorr}.
We can use the OPE to decompose the stress-tensor two-point function 
on $S^1_\beta\times \mathbb{R}^3$:
\begin{equation}\label{eq:ConfBlockExp}
\hat{G}_{\mu\nu,\rho\sigma}=	\langle T_{\mu\nu}(x)T_{\sigma\rho}(0)\rangle_\beta  = {1\over |x|^8}\sum_{\Delta,J,i_{n_J}}  \rho_{{\cal O},i} ~ g^{(i)}_{\Delta,J,\mu\nu,\rho\sigma}(x^\mu),  ~~~ \rho_{{\cal O},i} = \lambda_{TT{\cal O}}^{(i)}\langle {\cal O}\rangle_\beta,
\end{equation}
where we sum over operators in the $T\times T$ OPE and $i$ labels the different structures in the  OPE. 
For further details on the conformal blocks, 
see Appendix \ref{App:CFT} and also Appendix C in \cite{Karlsson:2022osn}. 
Integrating over the $xy$-plane, we will compare the OPE \eqref{eq:ConfBlockExp} against the bulk results in Section \ref{Sec:Bulk}.

We consider the OPE up to ${\cal O}(({x\over \beta})^8)$. The operators that contribute are the identity operator, the stress-tensor operator, and the double-stress tensors $[T^2]_J$ of the schematic form $:T_{\alpha\beta}T^{\alpha\beta}:$~, $:T_{\mu\alpha}{T^{\alpha}}_{\nu}:$~, and  $:T_{(\mu\nu}T_{\rho\sigma)}:$~, with spin $J=0,2,4$, respectively.  
For the double-stress tensors, we denote 
\begin{equation}
\rho_{i,J} = \rho_{[T^2]_{J,i}}
\end{equation}
where $i=\{1\}$ for $J=0$, $i=\{1,2\}$ for $J=2$, and $i=\{1,2,3\}$ for $J=4$. Perturbatively in $C_T^{-1}$, the coefficients $\rho_{i,J}$ and the anomalous dimensions $\Delta_J=8+\gamma_J$ are given by
\begin{equation}\label{eq:dataExp}
	\begin{aligned}
		\rho_{i,J} = \rho^{(0)}_{i,J}\Big[1+{\rho^{(1)}_{i,J}\over C_T}+\ldots\Big] \ , ~~~~~ \Delta_{J} = 8+{\gamma^{(1)}_J\over C_T}+\ldots \ .
	\end{aligned}
\end{equation}
The leading terms $\rho^{(0)}_{i,j}\sim C_T^2$ are due to the disconnected contribution to $\langle TT[T^2]_J\rangle$. This, in turn, produces the factorized part of the stress-tensor two-point function. Namely, to leading order in $C_T$, the correlator reads 
\begin{equation}\label{eq:Factorization}
	\langle T_{\mu\nu}(x)T_{\rho\sigma}(0)\rangle_\beta = \langle T_{\mu\nu}\rangle_\beta\langle T_{\rho\sigma}\rangle_\beta+{\cal O}(C_T) \ .
\end{equation}
Imposing factorization \eqref{eq:Factorization} fixes $5$ out of $6$ coefficients $\rho^{(0)}_{i,J}$ \cite{Karlsson:2022osn}:
\begin{equation}\label{eq:MFT}
	\begin{aligned}
		\rho_{1,2}^{(0)} = {324\over 7}\rho_{1,0}^{(0)} \ , ~~~ & \rho_{2,2}^{(0)} = {-1728\over 7}\rho_{1,0}^{(0)} \ ,\cr
		\rho_{1,4}^{(0)} = {160\over 7}\rho_{1,0}^{(0)}\ ,~~~ &\rho_{2,4}^{(0)} = {-1760\over 7}\rho_{1,0}^{(0)} \ , ~~~ \rho_{3,4}^{(0)} = {-480\over 7}\rho_{1,0}^{(0)} \ . \cr
	\end{aligned}
\end{equation}
The remaining coefficient $\rho_{1,0}^{(0)}$ is fixed by the non-zero diagonal terms in \eqref{eq:Factorization}. 

The thermal one-point function of a symmetric traceless operator $\OO$ on $S_\beta^1\times \mathbb{R}^3$ is fixed by symmetry up to a coefficient  $b_{\OO}$ (see, $e.g.$,  \cite{El-Showk:2011yvt,Iliesiu:2018fao})
\begin{equation}
	\langle \OO_{\mu_1\ldots\mu_J}\rangle_\beta = \frac{b_\OO}{\beta^{\Delta_\OO}}(e_{\mu_1}\ldots e_{\mu_J}-(traces)) \ ,
\end{equation}
where $e_\mu$ is a unit vector along the $S^1_\beta$.  In particular, by the thermalization of the stress tensor in a heavy state with $\Delta_H\sim C_T$ we have  
\begin{equation}
	\langle T_{\mu\nu}\rangle_\beta \approx \langle T_{\mu\nu}\rangle_H \ ,
\end{equation}
from which we find 
\begin{equation}
	\frac{b_{T_{\mu\nu}} }{\beta^4}= -\frac{d\Delta_H}{(d-1)S_4} \ ,
\end{equation}
where on the RHS we have inserted the OPE coefficient and $S_d=\frac{2 \pi^{d\over 2}}{\Gamma(d/2)}$. The relation between $\Delta_H$ and the parameter $\tilde{\mu}$ is given in \eqref{eq:defMuTilde} which leads to the relation:
\begin{equation}\label{eq:oneptT}
	\frac{b_{T_{\mu\nu}}}{\beta^4} = -\frac{C_T S_4(1+\kappa)^3\tilde{\mu}}{320\kappa} \ .
\end{equation}
Furthermore, plugging the MFT solution \eqref{eq:MFT} into the conformal block decomposition \eqref{eq:ConfBlockExp} together with factorization, one finds (to leading order in $C_T^{-1}$) 
\begin{equation}
	\langle T_{tt}\rangle_\beta^2 = 675 \rho^{(0)}_{1,0}  \ .
\end{equation} 
Inserting the stress-tensor one-point function in terms of $\tilde{\mu}$ from \eqref{eq:oneptT} gives 
\begin{equation}\label{eq:resOnePt}
	\rho^{(0)}_{1,0} = \frac{\pi^4 C_T^2(1+\kappa)^6\tilde{\mu}^2}{30720000\kappa^2} \ .
\end{equation}

\subsection{Stress-Tensor Contribution}

We first consider the stress-tensor contribution in the $T\times T$ OPE to the thermal two-point function. The stress-tensor three-point function is fixed by conformal symmetry up to three OPE coefficients $(\hat{a},\hat{b},\hat{c})$ in $d=4$ \cite{Osborn:1993cr} and the contribution to the stress-tensor two-point function at finite temperature was studied in, $e.g.$,  \cite{Kulaxizi:2010jt,Karlsson:2022osn}.\footnote{In particular, the contribution to the stress-tensor two-point functions $\hat{G}_{xy,xy}$, $\hat{G}_{tx,tx}$ and $\hat{G}_{tz,tz}$ can be found in Eq.\ (C.24) and (C.27) in \cite{Karlsson:2022osn} which, after integrating over the $xy$-plane, is given by Eq.\ (C.25), (C.28) and (C.30) in the same paper.  We shall not repeat them here due to their lengthy and unilluminating form.} 

Here, we are interested in the values for $(\hat{a},\hat{b},\hat{c})$ computed holographically in Gauss-Bonnet gravity. 
It was found in \cite{Buchel:2009sk} that  
\begin{equation}\label{eq:t2GB}
	t_{2,GB} = \frac{4f_\infty \lambda_{GB}}{1-2 f_\infty \lambda_{GB}}\frac{d(d-1)}{(d-2)(d-3)} \ , ~~~~ t_{4,GB}=0
\end{equation}
with the remaining coefficient fixed by Ward identities. 
The relation to the $(\hat{a},\hat{b},\hat{c})$ and $(t_2, t_4, C_T)$ bases can be found in \eqref{eq:abcGB}. 
We will be interested in the conformal collider bounds \cite{Hofman:2008ar}:
\begin{equation}
	\begin{aligned}
		(1-\frac{t_2}{3}-\frac{2t_4}{15})\geq 0 \ , ~~~~ 2(1-\frac{t_2}{3}-\frac{2t_4}{15})+t_2\geq 0 \ , ~~~~ \frac{3}{2}(1-\frac{t_2}{3}-\frac{2t_4}{15})+t_2+t_4\geq 0 \ , 
	\end{aligned}
\end{equation}
which for $t_2=t_{2,GB}$ and $t_4=0$ reduce to 
\begin{equation}\label{eq:ccbGB}
	\begin{aligned}
              (\kappa-\frac{4}{5})\geq 0 \ , ~~~~~ (2-\kappa)\geq 0 \ , ~~~~~ (\frac{4}{3}-\kappa)\geq 0 \ , 
	\end{aligned}
\end{equation}
where $\kappa=\sqrt{1-4\lambda_{GB}}$. 
The bounds are saturated for $\kappa=\{\frac{4}{5},2,\frac{4}{3}\}$. 

In \cite{Kulaxizi:2010jt}, the stress-tensor two-point function at finite temperature in the OPE expansion was considered in momentum space. In particular, the leading term in the lightcone limit due the stress-tensor contribution in the OPE was proportional to the conformal collider bounds in the respective channel. We now study this in position space after integrating over the $xy$-plane in the context of Gauss-Bonnet gravity. 

Using \eqref{eq:defMuTilde} together with $(\hat{a},\hat{b},\hat{c})$ \eqref{eq:abcGB} relevant for Gauss-Bonnet gravity, we find\footnote{The corresponding results  in terms of $(\Delta_H,\hat{a},\hat{b},\hat{c})$
can be found  in  Eq.\ (C.25), (C.28) and (C.30) in \cite{Karlsson:2022osn}.} 
\begin{equation}
	\begin{aligned}\label{eq:ResultStressTensor}
		G_{xy,xy}|_{\tilde{\mu}} &= \frac{(5\kappa -4) \pi  C_T (1+\kappa)^3\tilde{\mu} (t^2-z^2)
			}{800 \kappa ^2
			\left(t^2+z^2\right)^2} \ , \\
              G_{tx,tx}|_{\tilde{\mu}} &= -\frac{\pi  C_T(1+\kappa)^3\tilde{\mu}  \left((13 \kappa -4)
			t^4+6 (\kappa -2) t^2 z^2+(8-15 \kappa )
			z^4\right)}{6400 \kappa ^2 
			\left(t^2+z^2\right)^3} \ , \\
		G_{tz,tz}|_{\tilde{\mu}} &= \frac{\pi  C_T(1+\kappa)^3\tilde{\mu}  \left(-21 (3 \kappa +2)
			t^6+3 (94-93 \kappa ) t^4 z^2+(39 \kappa +98) t^2
			z^4+(111 \kappa -34) z^6\right)}{28800 \kappa ^2
		\left(t^2+z^2\right)^4} \ . \\
	\end{aligned}
\end{equation}
The result for $G_{xy,xy}$ in \eqref{eq:ResultStressTensor} is in agreement with the bulk computation in \eqref{mu1res}. To compare the remaining two polarizations with the bulk results, we apply the differential operators ${\cal D}^{2p}=(\partial_t^2+\partial_z^2)^p$ with $p=1$ for $G_{tx,tx}$ and $p=2$ for $G_{tz,tz}$. The results are
\begin{equation}
	\begin{aligned}
		{\cal D}^2G_{tx,tx}|_{\tilde{\mu}} &= -\frac{ (\kappa -2) 3 \pi  C_T(1+\kappa)^3 \tilde{\mu}
			\left(t^4-6 t^2
			z^2+z^4\right)}{1600 \kappa ^2
			 \left(t^2+z^2\right)^4} \ ,  \cr 
		{\cal D}^4G_{tz,tz}|_{\tilde{\mu}} &= \frac{(3 \kappa -4)  \pi  C_T(1+\kappa)^3 \tilde{\mu} 
			\left(t^6-15 t^4 z^2+15 t^2
			z^4-z^6\right)}{30 \kappa ^2
			\left(t^2+z^2\right)^6} \ ,
	\end{aligned}
\end{equation}
which agree with the bulk results in \eqref{eq:Bulktxtx1} and \eqref{tzmu1}. It follows that when $\kappa=\{\frac{4}{5},2,\frac{4}{3}\}$, the stress-tensor contribution to $G_{xy,xy}$, ${\cal D}^2 G_{tx,tx}$ and ${\cal D}^4 G_{tz,tz}$ vanishes.

\subsection{Double-Stress Tensor Contributions}

In the previous section, we saw that when a conformal collider bound is saturated, the contribution due to the stress-tensor operator to the $TT$ correlators at finite temperature vanishes for the corresponding polarization. 
In the lightcone limit at $\OO(({x\over \beta})^{4 k})$, the only operator that contributes is the multi-stress tensor operators on the leading Regge trajectory $[T^k]_{\mu_1\mu_2\ldots\mu_{2k}}$ (with spin $J=2k$). The bulk computation shows that not only does the stress-tensor contribution vanish when the conformal collider bounds are saturated, but the full contribution from the leading Regge trajectory also vanishes for the same choice of polarization. 

Below, we will read off the conformal data of double-stress tensors by comparison to the bulk results in Section \ref{Sec:Bulk}, following closely \cite{Karlsson:2022osn}. 
With this, we will see how the leading terms in the lightcone limit vanish when ANECs are saturated, which we further relate to the saturation of higher-spin ANECs in the next section.

We now consider the contribution due to double-stress tensors $[T^2]_J$ with $J=0,2,4$ to $G_{\mu\nu,\rho\sigma}$. 
We again use the OPE \eqref{eq:ConfBlockExp} and expand the dynamical data \eqref{eq:dataExp} to subleading order in $C_T$. 
The disconnected contribution was discussed above which gave the MFT coefficients $\rho_{i,J}^{(0)}$ in \eqref{eq:MFT} and \eqref{eq:resOnePt}.\footnote{The expression for the integrated conformal blocks expanded to subleading order in $C_T^{-1}$ can be found in Appendix C.5 in \cite{Karlsson:2022osn}. Only the overall normalization differs due to different values for $\rho^{(0)}_{1,0}$.}
Note that we need to regulate the integrals over the $xy$-plane which we do by inserting a factor of $(t^2+x^2+y^2+z^2)^{-{\epsilon\over 2}}$. As in \cite{Karlsson:2022osn}, we determine the double-stress tensor CFT data by imposing that the conformal block decomposition in terms of the CFT data agrees with the bulk results obtained in Section \ref{Sec:Bulk}:
\begin{equation}
	\begin{aligned}
		G_{xy,xy}^{(CFT)}-G_{xy,xy}^{(bulk)}\Big|_{\tilde{\mu}^2 C_T} = 0 \ , \\
		{\cal D}^2\Big[G_{tx,tx}^{(CFT)}-G_{tx,tx}^{(bulk)}\Big]\Big|_{\tilde{\mu}^2 C_T} = 0 \ , \\
		{\cal D}^4\Big[G_{tz,tz}^{(CFT)}-G_{tz,tz}^{(bulk)}\Big]\Big|_{\tilde{\mu}^2 C_T} = 0 \ . 
	\end{aligned}
\end{equation}

Using the bulk results \eqref{Gxy2}, \eqref{eq:Bulktxtx2} and \eqref{tzmu2} together with the conformal block expansion \eqref{eq:ConfBlockExp}, we find
\begin{equation}\label{eq:anomDimGB}
	\begin{aligned}
		\gamma^{(1)}_0 &= -\frac{80 \left(2103 \kappa ^2-4464 \kappa +2392\right)}{63 \pi ^4 \kappa ^2} \ ,\cr
		\gamma^{(1)}_2&= \frac{10 \left(19563 \kappa ^2-39996 \kappa +20012\right)}{189 \pi ^4 \kappa ^2}\ , \cr
		\gamma^{(1)}_4 &= -\frac{2 \left(24157 \kappa ^2-51412 \kappa +30228\right)}{105 \pi ^4 \kappa ^2},
	\end{aligned}
\end{equation}
and 
\begin{equation}\label{eq:OPEGB}
	\begin{aligned}
		&\rho_{2,2}^{(1)} = \frac{5 \left(157699 \kappa ^2-323228 \kappa +162636\right)}{1296 \pi ^4 \kappa ^2}+\rho _{1,2}^{(1)}  \ ,\cr
		&\rho_{2,4}^{(1)} =\frac{108521 \kappa ^2-170036 \kappa +65684}{2310 \pi ^4 \kappa ^2}+\rho _{1,4}^{(1)} \ ,\cr
		&\rho_{3,4}^{(1)} = \frac{-4053 \kappa ^2-14652 \kappa +21788}{1260 \pi ^4 \kappa ^2}+\rho _{1,4}^{(1)}\ ,
	\end{aligned}
\end{equation}
which reduce to the pure Einstein gravity results in \cite{Karlsson:2022osn} when $\kappa=1$. The remaining coefficients $(\rho_{1,0}^{(1)},\rho_{1,2}^{(1)},\rho_{1,4}^{(1)})$ are undetermined in the near-boundary analysis in the bulk, as mentioned in Section \ref{Sec:Bulk}.

Consider now the lightcone limit $(x^+,x^-)=(it+z,it-z)$ with $x^-\to 0$. Doing so, we find 
\begin{equation}\label{eq:TsqLC}
	\begin{aligned}
		G_{xy,xy}^{(CFT)}(x^+,x^-)\Big|_{\tilde{\mu}^2C_T} &\underset{x^-\to 0}{=} -\frac{ (4-5 \kappa )^2  \pi (1+\kappa)^6 C_T\tilde{\mu }^2}{153600 \kappa ^4 }\frac{(x^+)^3}{x^-} \ , \cr
		G_{tx,tx}^{(CFT)}(x^+,x^-)\Big|_{\tilde{\mu}^2C_T} &\underset{x^-\to 0}{=} -\frac{(\kappa -2)^2 17 \pi(1+\kappa)^6   C_T  \tilde{\mu }^2}{68812800 \kappa ^4  }\frac{(x^+)^4}{(x^-)^2} \ , \cr
		G_{tz,tz}^{(CFT)}(x^+,x^-)\Big|_{\tilde{\mu}^2C_T} &\underset{x^-\to 0}{=} -\frac{(4-3 \kappa )^2  11 \pi  (1+\kappa)^6 C_T \tilde{\mu }^2}{387072000 \kappa ^4 }\frac{(x^+)^5}{(x^-)^3} \ , 
	\end{aligned}
\end{equation}
where we note that this contribution comes solely from the spin-$4$ operator.\footnote{This property can be seen in Eq.\ (3.33) in \cite{Karlsson:2022osn}.}
Moreover, the near-lightcone behaviour is completely determined by the data in \eqref{eq:anomDimGB} and \eqref{eq:OPEGB}.  

In the lightcone limit,  when the conformal collider bounds are saturated, $i.e.$, $\kappa=\{\frac{4}{5},2,\frac{4}{3}\}$, both the stress-tensor and the spin-$4$ double-stress tensor contributions vanish.  As we will see in the following section, this is related to the saturation of the spin-$4$ ANEC, where the spin-$4$ operator is the double-stress tensor of the schematic form $:T_{(\mu\nu}T_{\rho\sigma)}:$.

\section{ANEC Interference Effects and Spin-4 ANEC}\label{ANECsection}

In this section, we study interference effects of the ANEC as well as the spin-4 ANEC. Interference effects in large-$C_T$ CFTs impose strong constraints on the MFT OPE coefficients.  We will see explicitly that the MFT OPE coefficients for the double-stress tensors, \eqref{eq:MFT}, are consistent with interference effects. 
In particular, we  verify that when the spin-$2$ ANEC is saturated
the spin–$4$ ANEC, the null-integrated $[T^2]_{J=4}$ double-stress tensor in holographic Gauss-Bonnet gravity, is also saturated in a stress-tensor state. 

Assuming a holographic CFT with a large $C_T$ and no light scalars, 
 the leading Regge trajectory of the $d=4$ stress-tensor OPE takes the following schematic form:
\begin{equation}
	T(x)T(0) = x^{-8}\Big[1+x^4 T(0)+ x^{\Delta_{[T^2]_4}}[T^2]_{J=4}(0)+\ldots\Big],
\end{equation}
where the ellipses denote higher-spin operators on the leading Regge trajectory, $i.e.$, multi-stress tensors $[T^k]_{J=2k}$ as well as all other operators. 
When integrated over a light-ray, the operators on the leading Regge trajectories $\OO^{(J)}$ are positive operators, see, $e.g.$,\ \cite{Komargodski:2016gci,Hartman:2016lgu,Meltzer:2017rtf,Meltzer:2018tnm}:
\begin{equation}
	\EE^{(J)} = \int_{-\infty}^\infty dx^- \OO^{(J)}_{-,-,\ldots,-}(x^-,0) \ , ~~~~ J=2,4,6, \ldots \ .
\end{equation}
In putative holographic CFTs dual to pure gravity 
in the bulk, the operators on the leading Regge trajectory are the multi-stress tensors $\OO^{(J)} = [T^k]_{J=2k}$. 
These are the ones that we will study. 
In particular, by studying matrix elements of $\EE^{(J)}$ in states that are superpositions of the stress tensor and multi-stress tensors, the positivity of the ANEC and higher-spin ANECs impose constraints on the stress-tensor OPE. 

To begin with, we consider the ANEC $\EE^{(2)}>0$ following \cite{Cordova:2017zej,Meltzer:2017rtf,Meltzer:2018tnm} and verify that it is satisfied in states of the schematic form $|\psi_J\rangle = v_1|T\rangle + v_2|[T^2]_{J}\rangle$ with $J=0,2,4$. 
This leads to a positive definite matrix schematically given by 
\begin{equation}\label{eq:ANEC}
	\langle\psi_J| \EE^{(2)}|\psi_J\rangle^{(i)} = v^\dagger \begin{pmatrix}
		\langle T|\EE^{(2)}|T\rangle & \langle T| \EE^{(2)}|[T^2]_J\rangle \\
		\langle [T^2]_J| \EE^{(2)}|T\rangle &\langle [T^2]_J| \EE^{(2)}|[T^2]_J\rangle
	\end{pmatrix}^{(i)}v\geq  0\,
\end{equation}
where the superscript $(i)$ labels different structures. Note that the entries are in general matrices. One then obtains bounds of the schematic form:  
\begin{equation}\label{eq:boundsSpin2ANEC}
	f^{(i)}(\{\Delta\},\{J\}) (\langle TT[T^2]_J\rangle^{(i)})^2 \leq \langle [T^2]_JT[T^2]_J\rangle^{(i)} \langle TTT\rangle^{(i)} 
\end{equation}
where $f^{(i)}(\{\Delta\},\{J\})$ is some function which depends on the scaling dimensions, spins,  
and the kinematical structure independent of the details of a theory. 

We expect that \eqref{eq:ANEC} in holographic CFTs 
has a $C_T$ scaling like follows
\begin{equation}
	v^\dagger\begin{pmatrix}
		\langle \tilde{T}| \EE^{(2)}|\tilde{T}\rangle & \langle \tilde{T}| \EE^{(2)}|[\tilde{T}^2]_J\rangle \\
		\langle [\tilde{T}^2]_J| \EE^{(2)}|\tilde{T}\rangle &\langle [\tilde{T}^2]_J| \EE^{(2)}|[\tilde{T}^2]_J\rangle
	\end{pmatrix}^{(i)}v = v^\dagger\begin{pmatrix}
		m_1 & C_T^{1/2} m_2\\
		C_T^{1/2} m_3 & m_4
	\end{pmatrix}^{(i)}v\geq 0 \ ,
\end{equation}  
for some $\OO(1)$ matrices $m_i$.  Here $\tilde{T}$ and $\tilde{T}^2$ denote unit-normalized operators/states.
By an appropriate choice of $v$, the above matrix requires positivity of any $2\times2$ submatrix. By a suitable choice of $v$, one can obtain terms of $\OO(C_T^{1/2})$ from the off-diagonal part and $\OO(1)$ terms from the diagonal part; this leads to potential positivity violations. Below, we will explicitly examine the spin-$2$ ANEC in the states $|\psi_J\rangle$ and show that the solution \eqref{eq:MFT} is consistent with positivity.

In what follows, we use the following three-point function basis \cite{Costa:2011mg,Costa:2011dw}:
{\small
\begin{equation}\label{eq:spinningBasis}
	\langle \OO_{\Delta_1,J_1} \OO_{\Delta_2,J_2} \OO_{\Delta_3,J_3}\rangle = \sum_{n_{12},n_{13}n_{23}} c^{(123)}_{n_{23},n_{13},n_{12}}{V_1^{J_1-n_{12}-n_{13}} V_2^{J_2-n_{23}-n_{12}} V_3^{J_3-n_{13}-n_{23}} H_{12}^{n_{12}}H_{13}^{n_{13}}H_{23}^{n_{23}} \over x_{12}^{\beta_{123}} x_{13}^{\beta_{132}} x_{23}^{\beta_{231}}},
\end{equation}}with $\beta_{ijk}=\beta_i+\beta_j-\beta_k$ and $\beta_i=\Delta_i+J_i$.   
This notation will be convenient to compare the data in the differential basis used in this work with the results of 
\cite{Meltzer:2017rtf,Meltzer:2018tnm}.

\subsection{Spin-0 Double-Stress Tensor Interference}

Interference effects between the stress-tensor state and a scalar was considered in \cite{Cordova:2017zej} which found that the function $f(\Delta)$ appearing in \eqref{eq:boundsSpin2ANEC} has (double) zeroes at $\Delta=2d+n$, 
where $\Delta$ refers to the dimension of the $T^2$ operator in $|\psi_0\rangle$.   
Due to the double-zero, there's no violation of the ANEC when considering interference effects in the state $|\psi_0\rangle$ to leading order in $C_{T}^{-1}$.

\subsection{Spin-2 Double-Stress Tensor Interference}

In \cite{Meltzer:2017rtf}, the positivity of the ANEC operator in a mixed state of a stress tensor and a spin-$2$ operator was studied. To this end, consider the state 
\begin{equation}
	|\psi_2\rangle = v_1|T\rangle +v_2 |[T^2]_{2}\rangle \ . 
\end{equation}
Due to the large-$C_T$ expansion, there is again a potential issue with the ANEC for the mixed stress tensor and spin-2 double-stress tensor state.  
It was explained in \cite{Meltzer:2017rtf} that
 if one parameterizes the three-point function $\langle TT[T^2]_2\rangle $ by $c^{(T[T^2]_2T)}_{0,0,0}$ and $c^{(T[T^2]_2T)}_{1,0,1}$ in the basis \eqref{eq:spinningBasis}, 
and imposes conservation, the ANEC positivity implies that 
$c^{(T[T^2]_2T)}_{0,0,0}=0$ while $c^{(T[T^2]_2T)}_{1,0,1}$ is unconstrained. 

Translating between the basis $(c^{(T[T^2]_2T)}_{0,0,0},c^{(T[T^2]_2T)}_{1,0,1})$ and the differential basis$(\rho_{1,2}^{(0)},\rho_{2,2}^{(0)})$, we find\footnote{Note that coefficients $\rho$ is a product of OPE coefficients and the thermal one-point function.} 
\begin{equation}\label{eq:spin2Coeffs}
	\begin{aligned}
		c^{(T[T^2]_2T)}_{0,0,0} = -\frac{96}{7} \left(16 \rho_{1,2}^{(0)}+3 \rho_{2,2}^{(0)}\right) \ , ~~~ c^{(T[T^2]_2T)}_{1,0,1} = \frac{1}{63} \left(2108 \rho_{1,2}^{(0)}+89\rho_{2,2}^{(0)}\right). 
	\end{aligned}
\end{equation} The superscript denotes the leading $C_T$ expressions, corresponding to $\Delta_{T^2}=8$ in $d=4$. 
Inserting the MFT solution \eqref{eq:MFT} in \eqref{eq:spin2Coeffs} gives
\begin{equation}
	\begin{aligned}
		c^{(T[T^2]_2T)}_{0,0,0} = 0 \ , ~~~ c^{(T[T^2]_2T)}_{1,0,1} =1200\rho_{1,0}^{(0)} \ .  
	\end{aligned}
\end{equation}
We see that $c^{(T[T^2]_2T)}_{0,0,0}=0$ while $c^{(T[T^2]_2T)}_{1,0,1}$ is unconstrained, showing consistency with the ANEC to leading order in $C_T^{-1}$ as discussed in \cite{Meltzer:2017rtf}.

\subsection{Spin-4 Double-Stress Tensor Interference}

Interference effects of both the ANEC and the spin-4 ANEC was studied in \cite{Meltzer:2018tnm}. 
There is again a potential issue with off-diagonal term that gives the leading large $C_T$ contribution when the minimal-twist spin-$4$ operator has dimension $\Delta=8+\OO(C_T^{-1})$ in holographic CFTs. This potentially leads to violations of the ANEC, but we will show that this is not the case based on the solution \eqref{eq:MFT}. 

Below, we define $\Theta=[T^2]_4$ and also denote the matrix elements of $\langle \OO_1|\EE^{(2)}|\OO_2\rangle^{(j)}$ by $\EE^{(2,j)}_{\OO_1\OO_2}$. 
Based on the results obtained in  \cite{Meltzer:2018tnm}, we obtain\footnote{More precisely, we take Eq.\ (C.9) in \cite{Meltzer:2018tnm} to obtain $\mathcal{E}^{(2,i)}_{T\Theta}$ in terms of $\mathcal{E}^{(4,j)}_{TT}$  and then use Eq.\ (C.2)-(C.4) in \cite{Meltzer:2018tnm} to express $\mathcal{E}^{(2,i)}_{T\Theta}$ in terms of the OPE coefficients $(c_{0,0,2}^{(T\Theta T)},c_{0,1,1}^{(T\Theta T)},c_{1,0,1}^{(T\Theta T)})$ for the basis \eqref{eq:spinningBasis}.  We refer the reader to \cite{Meltzer:2018tnm} for more details.} 
\begin{equation}
	\begin{aligned}
		\mathcal{E}^{(2,0)}_{T\Theta} &= \frac{1053\mathcal{E}^{(4,0)}_{TT} +748 \mathcal{E}^{(4,1)}_{TT}+128 \mathcal{E}^{(4,2)}_{TT}}{2419200} \ ,\cr
		\mathcal{E}^{(2,1)}_{T\Theta} &= \frac{319\mathcal{E}^{(4,0)}_{TT} +1284 \mathcal{E}^{(4,1)}_{TT}+204 \mathcal{E}^{(4,2)}_{TT}}{3225600} \ ,\cr
		\mathcal{E}^{(2,2)}_{T\Theta} &= \frac{217\mathcal{E}^{(4,0)}_{TT} +852 \mathcal{E}^{(4,1)}_{TT}+1752 \mathcal{E}^{(4,2)}_{TT}}{9676800} \ .
	\end{aligned}
\end{equation}
Due to the large-$C_T$ scaling, we need to impose $\mathcal{E}^{(2,i)}_{T\Theta }=0$ to leading order in $C_T^{-1}$; otherwise we would find violations of the ANEC. However, each $\mathcal{E}^{(4,i)}_{TT}$ is non-negative which implies that $\mathcal{E}^{(4,i)}_{TT}=0$  to leading order in $C_T$. In terms of $(c_{0,0,2}^{(T\Theta T)},c_{0,1,1}^{(T\Theta T)},c_{1,0,1}^{(T\Theta T)})$, we find the only solution is 
\begin{equation}
	\begin{aligned}
		c_{0,0,2}^{(T \Theta  T)}=0 \ , ~~~ c_{0,1,1}^{(T \Theta  T)}=0 \ , ~~~ c_{1,0,1}^{(T \Theta  T)}=0 \ , ~~~
	\end{aligned}
\end{equation}
which seems to imply that $\Theta=[T^2]_{J=4}$ cannot appear in the stress-tensor OPE.  
But this is not the case due to the behavior of the OPE coefficients as we now explain. 
Solving conservation and the permutation symmetry  in terms of the three coefficients $(c_{0,0,2}^{(T \Theta  T)},c_{0,1,1}^{(T \Theta  T)},c_{1,0,1}^{(T \Theta  T)})$, we find that all the coefficients are regular as $\Delta\to 8$ except for 
\begin{equation}
	c_{2,0,2}^{(T \Theta  T)} \sim {1\over\Delta-8}p(c_{0,0,2}^{(T \Theta  T)},c_{0,1,1}^{(T \Theta  T)},c_{1,0,1}^{(T \Theta  T)}) \ ,
\end{equation}
where $p(c_{0,0,2}^{(T \Theta  T)},c_{0,1,1}^{(T \Theta  T)},c_{1,0,1}^{(T \Theta  T)})$ is a linear function of the OPE coefficients $(c_{0,0,2}^{(T \Theta  T)},$ $c_{0,1,1}^{(T \Theta  T)},c_{1,0,1}^{(T \Theta  T)})$. 
Requiring that the three-point function is regular as $\Delta=\Delta_\Theta\to 8$, we write\footnote{Including the anomalous dimensions would lead to the coefficients having different scaling with $C_T$.} 
\begin{equation}
	\begin{aligned}
		\lim_{\Delta\to 8}c_{0,0,2}^{(T \Theta  T)} &= (\Delta-8)\tilde{c}_{0,0,2}^{(T \Theta  T)},\cr
		\lim_{\Delta\to 8}c_{0,1,1}^{(T \Theta  T)} &= (\Delta-8)\tilde{c}_{0,1,1}^{(T \Theta  T)},\cr
		\lim_{\Delta\to 8}c_{1,0,1}^{(T \Theta  T)} &= (\Delta-8)\tilde{c}_{1,0,1}^{(T \Theta  T)},
	\end{aligned}
\end{equation}
with constants $\tilde{c}$'s that are finite as $\Delta \to 8$. 
This does not imply that the three-point function is trivial due to the simple pole in $c_{2,0,2}^{(T \Theta  T)}$.  In particular, the three-point function is 
\begin{equation}\label{eq:attt}
	\langle T(P_1)[T^2]_4(P_2)T(P_3)\rangle = {\alpha H_{12}^2H_{23}^2\over (P_1\cdot P_2)^6(P_3\cdot P_2)^6} \ ,
\end{equation}
for some coefficient $\alpha$. As the three coefficients $(c_{0,0,2}^{(T \Theta  T)},c_{0,1,1}^{(T \Theta  T)},c_{1,0,1}^{(T \Theta  T)})$ all vanish as $\Delta\to 8$, the solution is consistent with $\EE_{T\Theta}^{(2,i)}=0$ to leading order in $C_T$.

Note that the leading Regge trajectory obey the inequalities 
\begin{equation}
	d-2\leq \tau_{J,min}<2(d-2) \ ,
\end{equation}
in interacting CFTs. Therefore including anomalous dimensions of $\OO(C_T^{-1})$ such that $\tau_4<4$, the coefficients  $(c_{0,0,2}^{(T \Theta  T)},c_{0,1,1}^{(T \Theta  T)},c_{1,0,1}^{(T \Theta  T)})$ can become non-zero and not violate the spin-$4$ interference effects.

We find that the solution \eqref{eq:attt} agrees with that of MFT \eqref{eq:MFT}. This can be seen by inserting \eqref{eq:MFT} into the explicit expressions for the three-point function in the differential basis, giving $\alpha={75\rho_{1,0}^{(0)}\over 26}$ in the three-point function \eqref{eq:attt}.\footnote{It can also be seen by solving for $c_{n_{23},n_{13},n_{12}}^{(T\Theta T)}=c_{n_{23},n_{13},n_{12}}^{(T \Theta  T)}(\rho_{1,4}^{(0)},\rho_{2,4}^{(0)},\rho_{3,4}^{(0)})$, from which one finds that all coefficients vanish except for $c_{2,0,2}^{(T \Theta  T)}$ (to leading order in $C_T^{-1}$).} 
Therefore, we conclude that the MFT coefficients \eqref{eq:MFT} are consistent with positivity of the ANEC in the state which is a superposition of the stress tensor and the spin-$4$ double-stress tensor $[T^2]_{J=4}$.

\subsection{Spin-4 ANEC in Stress-Tensor State}

We have seen how the MFT solution is consistent with the ANEC in states $|\psi_J\rangle$ that are superpositions of a stress-tensor and double-stress tensor state. We now move on to consider the spin-$4$ ANEC and study it when the  spin-$4$ operator is the double-stress tensor $[T^2]_{J=4}$ with the OPE data obtained in holographic Gauss-Bonnet theory. We will show that the saturation of the spin-$4$ ANEC happens precisely when the corresponding contribution to the near-lightcone $TT$ correlators at finite temperature vanishes, generalizing the results for the stress tensor in \cite{Kulaxizi:2010jt}. Note this analysis is sensitive to the subleading terms in the $C_T^{-1}$ expansion of the double-stress tensor data.

One can obtain the spin-4 ANEC in a stress-tensor state $\EE^{(4,j)}_{TT}$ using the results from \cite{Meltzer:2018tnm}.\footnote{See Eq.\ (4.4)-(4.6) in \cite{Meltzer:2018tnm}.} We change basis from $(c_{0,0,2}^{(T \Theta  T)},c_{0,1,1}^{(T \Theta  T)},c_{1,0,1}^{(T \Theta  T)})$ to the differential basis $(\rho_{1,4},\rho_{2,4},\rho_{3,4})$ used in the present  paper and perform the $C_T^{-1}$ expansion.\footnote{The results are proportional to the leading lightcone expressions in Eq.\ (3.33) in \cite{Karlsson:2022osn}.} 
Using the values in Gauss-Bonnet gravity given in \eqref{eq:anomDimGB} and \eqref{eq:OPEGB}, we obtain 
\begin{equation}\label{eq:spin4ANECGB}
	\begin{aligned}
		0\leq\EE^{(4,0)}_{TT} &=  \frac{11 \pi ^4 C_T(4-3 \kappa )^2 (\kappa +1)^6 \tilde{\mu }^2}{2211840000 \kappa ^4},\cr
		0\leq\EE^{(4,1)}_{TT} &=  \frac{17 \pi ^4 C_T(\kappa -2)^2 (\kappa +1)^6 \tilde{\mu }^2}{98304000 \kappa ^4},\cr
		0\leq\EE^{(4,2)}_{TT} &=  \frac{7 \pi ^4 C_T(4-5 \kappa )^2 (\kappa +1)^6 \tilde{\mu }^2}{3072000 \kappa ^4},
	\end{aligned}
\end{equation}
which saturates when $\kappa=\{{4\over 3},2,{4\over 5}\}$.

\section{Discussion}

In this paper, we study  thermal $TT$ correlators and explore their connections to  ANECs.
One can use the OPE between two stress tensors and expand the correlator in powers of the temperature.
The contributions from a single-stress tensor in the lightcone limit are proportional to
the corresponding spin-2 ANECs.
To go beyond it, we consider holographic Gauss-Bonnet gravity, where the breakdown of spin-$2$ ANECs is related to superluminal signal propagation.
We analyze the multi-stress tensor contributions to the $TT$ correlators in the dual $d=4$ CFT with a large central charge.
Our chief finding in this paper is that, when an  ANEC is saturated in a state created by the stress tensor, all higher-spin ANECs are saturated in this state as well -- the corresponding near-lightcone thermal $TT$ correlator takes the vacuum form.
 
 Note that the statement about ANEC saturation is really a statement about the OPE of
 the stress tensors, so instead of a thermal state one may consider any other suitable state in the theory.
 One may ask how general our observation is -- does it apply beyond holographic models
 and beyond the large $C_T$ limit?  Below, we discuss related questions and possible future directions.

\begin{itemize}

\item  {\it Scope of the result and possible proof:}

It was argued in \cite{Zhiboedov:2013opa,Meltzer:2018tnm}, that ANEC saturation implies that the theory is, in some sense, free. In particular, by studying ANECs in the states created by linear combinations of spin-2 and spin-4 operators, \cite{Meltzer:2018tnm} argued that the spin-4 operator must be a conserved current and hence the theory is free. However we found that things can be more subtle when the spin-4 operator has dimension eight, which is the case for the minimal-twist double-stress tensors in CFTs with a large $C_T$. In this case the theory is not free, and only thermal correlators with certain polarization simplify in the near-lightcone regime.

\qquad Are there  examples of unitary interacting CFTs which are ``free" near the lightcone, like holographic GB gravity we studied here? That would be an interesting question to investigate.  Once the scope of this phenomenon becomes more clear, it would be natural to search for a proof as well.

\item {\it Free theories and their large $N$ limit:} 

Free theories (bosons, fermions and gauge fields in four spacetime dimensions) saturate conformal collider bounds, so it is natural to ask what happens with the higher-spin ANECs in this case.
Of course, the near-lightcone behavior in free theories is governed by the conserved, higher-spin currents.
Nevertheless, it would be interesting to see if there are any patterns of the type we observed in this paper.
It seems that studying the large $C_T$ (or large $N$) limit of free theories might be particularly interesting;
we leave this for future work.

\item {\it Relation to experiment and to lattice computations:} 

One may wonder if there are CFTs which are interacting and at the same time saturate ANECs, like the holographic model we considered in this paper. Presumably a spin-four operator with conformal dimension close to eight might be necessary for this to happen. It would be interesting to check, how far, $e.g.$, QCD at finite temperature is from this regime and to compare our results with the lattice computations of the $TT$ correlators (see, $e.g.$, \cite{Meyer:2011gj} for a review).

\item  {\it Anomalous dimensions of the spin-2 $[T_{\mu\nu}]^2$ operator:}

We note that the anomalous dimension for the spin-$2$ double-stress tensor, given by the second equation in (\ref{eq:anomDimGB}), is negative for $\lambda_{GB}=0$ (Einstein gravity) but changes sign and becomes positive for values of $\lambda_{GB}$ inside the conformal collider bounds.  
It would be interesting to understand the meaning of these values of $\lambda_{GB}$ where this happens.

\item {\it Minimal-twist multi-stress tensors with derivatives and spherical black holes:} 

For technical reasons, in this paper we restrict our discussion to a black hole with a planar horizon. This corresponds to considering multi-stress tensor operators without additional derivatives appearing in the OPE.  
It would be interesting to study the role of operators with derivatives, although this would be technically more involved than the analysis we did in this paper.

\item  {\it Near-lightcone $TT$ correlators 
and 
higher-derivative gravities:}

On a related note, one may ask if one can make progress in computing the near-lightcone behavior of holographic correlators for generic holographic models.

\qquad Much recent progress has been made in understanding the multi-stress tensor sector of the $d=4$ thermal {\it scalar} two-point functions and related heavy-heavy-light-light (HHLL) correlators \cite{Fitzpatrick:2019zqz, Karlsson:2019qfi,Li:2019tpf,Huang:2019fog,Kulaxizi:2019tkd,Fitzpatrick:2019efk,Karlsson:2019dbd,Haehl:2019eae,Li:2019zba,Karlsson:2019txu,Huang:2020ycs,Karlsson:2020ghx,Li:2020dqm,Parnachev:2020fna,Fitzpatrick:2020yjb,Berenstein:2020vlp,Parnachev:2020zbr,Karlsson:2021duj,Rodriguez-Gomez:2021pfh,Huang:2021hye,Rodriguez-Gomez:2021mkk,Krishna:2021fus,Bianchi:2021yqs,Karlsson:2021mgg,Huang:2022vcs,Dodelson:2022eiz,Dodelson:2022yvn, Parisini:2022wkb}. 
As was observed in \cite{Huang:2021hye, Karlsson:2021mgg}, the structure of the $d=4$ thermal scalar two-point correlator in the lightcone limit has certain similarity with the ${\cal W}_3$ vacuum blocks in $d=2$ CFT \cite{Z1985}.  While the reasons for this remain unclear,  one may wonder whether a similar story exists for the $TT$ correlators.

\qquad For example, is there a universality of the near-lightcone $TT$ correlators similar to 
the one exhibited by the near-lightcone HHLL holographic correlators?
The addition of higher-derivative terms to the bulk gravitational Lagrangian leads to the variation of the $TTT$ couplings, but is the near-lightcone behavior of the holographic $TT$ correlators fixed (and universal) in terms of these couplings? Can the bootstrap techniques of \cite{Karlsson:2019dbd} be applied to compute the full $TT$ correlator? We leave these questions for future investigation.

\qquad Note that the model we consider, Gauss-Bonnet gravity,  can be regarded as the simplest type of the Lovelock theories \cite{lovelock1971}. We expect that the techniques used in our work can be used to deal with other higher-derivative corrections to the bulk Lagrangian.  Additional parameters present in such theories can also be useful for studying possible universality of the holographic thermal $TT$ correlators.

\item  {\it Finite-gap corrections:}

In the case of the stress-tensor sector of holographic HHLL correlators, the finite-gap corrections have been 
investigated in \cite{Fitzpatrick:2020yjb} and were shown to lead to the loss of universality.
It would be interesting to repeat this analysis for the $TT$ correlators.

\item {\it  Higher-point correlators:}  

Another natural extension of this work is to go further and investigate the thermal properties of $n$-point ($n>2$) stress-tensor correlators near the lightcone. 

\item {\it  Going beyond double-stress tensors:}  

In this paper, as well as in \cite{Karlsson:2022osn}, the conformal block decomposition of the holographic thermal $TT$ correlators has been performed up to the double-stress tensors.  It would be interesting to go beyond this and study the $k$-stress tensor contributions for generic values of $k$.

\end{itemize}

\section*{Acknowledgments}

We would like to thank  Manuela Kulaxizi, David Kutasov, David Meltzer  and Sasha Zhiboedov for discussions, correspondence 
and comments on the draft. 
KWH was supported in part by the Irish Research Council Government of Ireland Postdoctoral Fellowship
under project award number GOIPD/2022/288.
RK, AP, and SV were supported in part by an Irish Research Council consolidator award. 
RK 
was also supported by the European Research Council (ERC) under the European Union’s Horizon 2020 research and innovation programme (grant agreement number 949077).
A.P. thanks the University of Chicago, where part of this work was completed, for hospitality.

\appendix
    \section{More Shear-Channel Results}\label{Appendixxzxz}
    
When the source $\hat{H}_{xz}$ is turned on,  using the method discussed in \ref{shearHoloss},  we find
{
	\allowdisplaybreaks
	\begin{align}
G_{xz,xz}^{(bulk)}\Big|_{\tilde{\mu}^0}=&\frac{1}{\partial_t^2+\partial_z^2}\frac{3 \pi   C_T \left(7 t^2-z^2\right)}{5 \left(t^2+z^2\right)^5}\ , \\
G_{xz,xz}^{(bulk)}\Big|_{\tilde{\mu}}=&(\kappa -2) \frac{1}{\partial_t^2+\partial_z^2}\frac{3 \pi    \tilde{\mu } C_T \left(t^4-6 t^2 z^2+z^4\right)}{100 \kappa ^2 (\kappa +1) L^8  \left(t^2+z^2\right)^4}\ , \\
	G_{xz,xz}^{(bulk)}\Big|_{\tilde{\mu}^2}=&-\frac{1}{\partial_t^2+\partial_z^2}\Bigg[\frac{\pi  \tilde{\mu }^2 C_T}{2100 \kappa ^4 (\kappa +1)^2 L^{16} \left(t^2+z^2\right)^3} \Big(6 (\kappa  (277 \kappa -700)\nonumber\\
	&+388) t^6+24 (\kappa  (161 \kappa -418)+230) t^4 z^2+6 (\kappa  (311 \kappa -836)\nonumber\\
    &+524) t^2 z^4+3 (\kappa  (277 \kappa -700)+388) \left(t^2+z^2\right)^3 \log
   \left(t^2+z^2\right)\nonumber\\
   &-16 (\kappa  (38 \kappa -119)+71) z^6\Big)+\frac{3}{5} \pi C_T  a^{8,1(xz)}_{8,0}\Bigg].\label{xzxzmu2hr}
\end{align} 
}In the Einstein-gravity limit, $\kappa\rightarrow1$, these results agree with  \cite{Karlsson:2022osn}. The $\tilde{\mu}$ contribution vanishes when $\kappa=2$, the critical value of the GB coupling for this channel.

Next consider the $\tilde{\mu}^2$ contribution. In the lightcone limit, we find 
    \begin{equation}
        \begin{split}
            G_{tx,tx}^{(bulk)}(x^+, x^-)\Big|_{\tilde{\mu}^2} \underset{x^- \to 0}{=}
&\,\,-\frac{1}{\partial_+ \partial_-}\bigg(\frac{ (\kappa -2)^2 17 \pi  C_T (x^+)^3 \tilde{\mu }^2}{33600 \kappa ^4 (\kappa +1)^2 L^{16} (x^-)^3}\\
            &-\frac{\pi  C_T \big(\kappa  (107 \kappa -340)+212\big) (x^+)^2 \tilde{\mu }^2}{11200 \kappa ^4 (\kappa +1)^2 L^{16} (x^-)^2}+{\cal O}\big({1\over x^{-}}\big)\bigg) \ .
        \end{split}
    \end{equation}
The leading-lightcone contribution vanishes at the critical $\kappa=2$.

\vspace{5mm}

\noindent{\it Reduced equation of motion:} 

With $\hat{H}_{xz}$ turned on, the corresponding reduced equation of motion is  
\begin{align}
\mu^2_{\rm{eff}({\rm shear})}\Theta_{2({\rm shear})} + \mu_{\rm{eff}({\rm shear})}\Theta_{1({\rm shear})}+ \Theta_{0({\rm shear})}= 0  \ , ~~~
\mu_{\rm{eff}({\rm shear})}=  \big(\k- 2 \big) \tilde\mu
\end{align}  
where $\Theta_{2({\rm shear})}$ is the same as the $\hat{H}_{tx}$ result:
{\footnotesize{\begin{align}
{ \Theta_{2({\rm shear})}\over 4  v^4  } =& 
\Big[w^2 \left(w^2 \del^4_w-38 w \del^3_w+591 \del^2_w\right)-4431 w \del_w +13440  \Big]Q_{\text{tot}}\ .
\end{align}}}
In this case, $\Theta_{1({\rm shear})}$ and $\Theta_{0({\rm shear})}$ are given by 
{\footnotesize{\begin{align}
{\Theta_{1({\rm shear})}\over 2  \kappa ^2 (\kappa +1) L^8 v^2}
=&\Big[ 2 \left(w^2-1\right) w^5   \del^3_w  -2 v w^6 \del_v \del^2_w     +  6  \left(9-5 w^2\right) w^4 \del^2_w + 34 v w^5 \del_v \del_w \nn\\
&~~~~~ +  6  \left(21 w^2-89\right) w^3 \del_w  -160  w^4 v \del_v +1920 w^2 \Big]\bar Q\nn
\end{align}}}
\vspace{-8mm}
{\footnotesize{\begin{align}
 &+\Big[ \left(w^4-1\right) w^4 \del^4_w    -2 \left(w^2-3\right) v w^5   \del_v \del^3_w + v^2 w^6  \del^2_v \del^2_w  -2  \left(6 w^4+5 w^2-19\right) w^3 \del^3_w \nn\\
& ~~~~~-17 v^2 w^5 \del^2_v \del_w  
+(31 w^2 -162 )  v w^4  \del_v \del^2_w
+  (-591 + 270 w^2 + 17 w^4) w^2 \del^2_w \nn\\
&~~~~~+ 80 v^2 w^4 \del^2_v 
+(1602 - 143 w^2)  v w^3  \del_v \del_w  
+ (4431 - 2670 w^2 + 335 w^4) w \del_w \nn\\
&~~~~~+ 80  (w^2 -72 ) v w^2 \del_v  - 640 (2 w^4- 15 w^2+21)\Big] Q_{\text{tot}}\ , \\
{\Theta_{0({\rm shear})}\over \kappa ^4 (\kappa +1)^2 L^{16} w^2}
=& \Big[ 2  ( w^2-1)^2 w^3 \del^3_w +  2  (w^2-1) v w^4\del_v \del^2_w -4 v^2 w^5 \del_w \del^2_v  \nn\\
& -(  26 w^4- 80 w^2+54 )  w^2 \del^2_w + 32 v^2 w^4 \del^2_v  - 2 (  7 w^2-17)v w^3  \del_v \del_w \nn\\
& +2  (61 w^4- 296 w^2  +267  )  w \del_w-160 v w^2 \del_v -64 (3 w^4 - 25 w^2+30  )\Big]\bar Q\nn
\end{align}
\vspace{-10mm}
\begin{align}
&+ \Big[ ( w^2-1)^2 w^4\del^4_w +  2 v^3 w^5 \del_w \del^3_v + 4  (w^2-1) v w^3 \del^3_w \del_v - ( 3 w^2-7)  v^2 w^4 \del^2_w\del^2_v \nn\\
&~~~  -12 (  w^4 - 3 w^2+2) w^3  \del^3_w  -16  w^4  v^3 \del^3_v  + ( 5 w^4 - 89 w^2 +108) v w^2  \del^2_w \del_v  +  ( 29 w^2 -119 )v^2 w^3 \del_w \del^2_v \nn\\
&~~~ + 3  ( 19 w^4- 82 w^2 +71)  w^2 \del^2_w -16 (4 w^2-35 )   w^2 v^2 \del^2_v - (55 w^4- 745 w^2+1068  ) v w \del_w \del_v \nn\\
&~~~ -21  ( 5 w^4-30 w^2
+33) w\del_w + 80 (2 w^4- 29 w^2+48  ) v \del_v  \Big] Q_{\text{tot}} \ . 
\end{align}}} 
\vspace{-8mm}

\section{Coefficients of the Stress-Tensor Three-Point Function}\label{App:CFT}

The stress-tensor three-point function is parameterized by three coefficients $(\hat{a},\hat{b},\hat{c})$ \cite{Osborn:1993cr}. 
An alternative basis uses $(t_2,t_4,C_T)$, which can be related to the previous basis in the following way \cite{Hofman:2008ar}: 
\begin{equation}
	\begin{aligned}
		t_2 = \frac{30(13\hat{a}+4\hat{b}-3\hat{c})}{14\hat{a}-2\hat{b}-5\hat{c}} \ , ~~~~ t_4 = -\frac{15(81\hat{a}+32\hat{b}-20\hat{c})}{2(14\hat{a}-2\hat{b}-5\hat{c})},
	\end{aligned}
\end{equation}
and \cite{Osborn:1993cr}
\begin{equation}
	C_T = 4S_d \frac{(d-2)(d+3)\hat{a}-2\hat{b}-(d+1)\hat{c}}{d(d+2)}.
\end{equation}  In this paper, we focus on $d=4$. The stress-tensor three-point function was studied in the context of  holographic Gauss-Bonnet gravity in \cite{Buchel:2009sk}, which found $(t_{2,GB},t_{4,GB},C_T)$. 
Setting $t_2=t_{2,GB}$ given in \eqref{eq:t2GB} and $t_{4,GB}=0$, one finds\footnote{One can obtain the stress-tensor contribution to the thermal $TT$ correlators using \eqref{eq:defMuTilde}, together with Eq.\ (C.25), (C.28) and (C.30) in  \cite{Karlsson:2022osn}.} 
\begin{equation}\label{eq:abcGB}
	\begin{aligned}
		\hat{a} = \frac{8 C_T \left(-6+\frac{5}{\k}\right)}{45 \pi ^2} \ , ~~~~
		\hat{b} = \frac{C_T \left(33-\frac{50}{\k}\right)}{90 \pi ^2} \ ,  ~~~~ 
            \hat{c} = \frac{2 C_T \left(-84+\frac{61}{\k}\right)}{45 \pi ^2}\ .
	\end{aligned}
\end{equation}

\bibliographystyle{JHEP}
\bibliography{GBrefs} 

\providecommand{\href}[2]{#2}\begingroup\raggedright\begin{thebibliography}{10}

\bibitem{Faulkner:2016mzt}
T.~Faulkner, R.~G. Leigh, O.~Parrikar, and H.~Wang, {\it {Modular Hamiltonians
  for Deformed Half-Spaces and the Averaged Null Energy Condition}},  {\em
  JHEP} {\bf 09} (2016) 038, [\href{http://arxiv.org/abs/1605.08072}{{\tt
  arXiv:1605.08072}}].

\bibitem{Hartman:2016lgu}
T.~Hartman, S.~Kundu, and A.~Tajdini, {\it {Averaged Null Energy Condition from
  Causality}},  {\em JHEP} {\bf 07} (2017) 066,
  [\href{http://arxiv.org/abs/1610.05308}{{\tt arXiv:1610.05308}}].

\bibitem{Hofman:2008ar}
D.~M. Hofman and J.~Maldacena, {\it {Conformal collider physics: Energy and
  charge correlations}},  {\em JHEP} {\bf 05} (2008) 012,
  [\href{http://arxiv.org/abs/0803.1467}{{\tt arXiv:0803.1467}}].

\bibitem{Rychkov:2016iqz}
S.~Rychkov, {\it {EPFL Lectures on Conformal Field Theory in D\ensuremath{>}= 3
  Dimensions}},  \href{http://arxiv.org/abs/1601.05000}{{\tt
  arXiv:1601.05000}}.

\bibitem{Simmons-Duffin:2016gjk}
D.~Simmons-Duffin, {\it {The Conformal Bootstrap}},  in {\em {Theoretical
  Advanced Study Institute in Elementary Particle Physics}: {New Frontiers in
  Fields and Strings}}, pp.~1--74, 2017.
\newblock \href{http://arxiv.org/abs/1602.07982}{{\tt arXiv:1602.07982}}.

\bibitem{Poland:2018epd}
D.~Poland, S.~Rychkov, and A.~Vichi, {\it {The Conformal Bootstrap: Theory,
  Numerical Techniques, and Applications}},  {\em Rev. Mod. Phys.} {\bf 91}
  (2019) 015002, [\href{http://arxiv.org/abs/1805.04405}{{\tt
  arXiv:1805.04405}}].

\bibitem{Hofman:2016awc}
D.~M. Hofman, D.~Li, D.~Meltzer, D.~Poland, and F.~Rejon-Barrera, {\it {A Proof
  of the Conformal Collider Bounds}},  {\em JHEP} {\bf 06} (2016) 111,
  [\href{http://arxiv.org/abs/1603.03771}{{\tt arXiv:1603.03771}}].

\bibitem{Li:2015itl}
D.~Li, D.~Meltzer, and D.~Poland, {\it {Conformal Collider Physics from the
  Lightcone Bootstrap}},  {\em JHEP} {\bf 02} (2016) 143,
  [\href{http://arxiv.org/abs/1511.08025}{{\tt arXiv:1511.08025}}].

\bibitem{Komargodski:2016gci}
Z.~Komargodski, M.~Kulaxizi, A.~Parnachev, and A.~Zhiboedov, {\it {Conformal
  Field Theories and Deep Inelastic Scattering}},  {\em Phys. Rev. D} {\bf 95}
  (2017), no.~6 065011, [\href{http://arxiv.org/abs/1601.05453}{{\tt
  arXiv:1601.05453}}].

\bibitem{Cordova:2017zej}
C.~C\'ordova, J.~Maldacena, and G.~J. Turiaci, {\it {Bounds on OPE Coefficients
  from Interference Effects in the Conformal Collider}},  {\em JHEP} {\bf 11}
  (2017) 032, [\href{http://arxiv.org/abs/1710.03199}{{\tt arXiv:1710.03199}}].

\bibitem{Meltzer:2017rtf}
D.~Meltzer and E.~Perlmutter, {\it {Beyond $a = c$: gravitational couplings to
  matter and the stress tensor OPE}},  {\em JHEP} {\bf 07} (2018) 157,
  [\href{http://arxiv.org/abs/1712.04861}{{\tt arXiv:1712.04861}}].

\bibitem{Meltzer:2018tnm}
D.~Meltzer, {\it {Higher Spin ANEC and the Space of CFTs}},  {\em JHEP} {\bf
  07} (2019) 001, [\href{http://arxiv.org/abs/1811.01913}{{\tt
  arXiv:1811.01913}}].

\bibitem{Balakrishnan:2017bjg}
S.~Balakrishnan, T.~Faulkner, Z.~U. Khandker, and H.~Wang, {\it {A General
  Proof of the Quantum Null Energy Condition}},  {\em JHEP} {\bf 09} (2019)
  020, [\href{http://arxiv.org/abs/1706.09432}{{\tt arXiv:1706.09432}}].

\bibitem{Cordova:2017dhq}
C.~C\'ordova and K.~Diab, {\it {Universal Bounds on Operator Dimensions from
  the Average Null Energy Condition}},  {\em JHEP} {\bf 02} (2018) 131,
  [\href{http://arxiv.org/abs/1712.01089}{{\tt arXiv:1712.01089}}].

\bibitem{Kravchuk:2018htv}
P.~Kravchuk and D.~Simmons-Duffin, {\it {Light-ray operators in conformal field
  theory}},  {\em JHEP} {\bf 11} (2018) 102,
  [\href{http://arxiv.org/abs/1805.00098}{{\tt arXiv:1805.00098}}].

\bibitem{Delacretaz:2018cfk}
L.~V. Delacr\'etaz, T.~Hartman, S.~A. Hartnoll, and A.~Lewkowycz, {\it
  {Thermalization, Viscosity and the Averaged Null Energy Condition}},  {\em
  JHEP} {\bf 10} (2018) 028, [\href{http://arxiv.org/abs/1805.04194}{{\tt
  arXiv:1805.04194}}].

\bibitem{Cordova:2018ygx}
C.~C\'ordova and S.-H. Shao, {\it {Light-ray Operators and the BMS Algebra}},
  {\em Phys. Rev. D} {\bf 98} (2018), no.~12 125015,
  [\href{http://arxiv.org/abs/1810.05706}{{\tt arXiv:1810.05706}}].

\bibitem{Ceyhan:2018zfg}
F.~Ceyhan and T.~Faulkner, {\it {Recovering the QNEC from the ANEC}},  {\em
  Commun. Math. Phys.} {\bf 377} (2020), no.~2 999--1045,
  [\href{http://arxiv.org/abs/1812.04683}{{\tt arXiv:1812.04683}}].

\bibitem{Belin:2019mnx}
A.~Belin, D.~M. Hofman, and G.~Mathys, {\it {Einstein gravity from ANEC
  correlators}},  {\em JHEP} {\bf 08} (2019) 032,
  [\href{http://arxiv.org/abs/1904.05892}{{\tt arXiv:1904.05892}}].

\bibitem{Kologlu:2019mfz}
M.~Kologlu, P.~Kravchuk, D.~Simmons-Duffin, and A.~Zhiboedov, {\it {The
  light-ray OPE and conformal colliders}},  {\em JHEP} {\bf 01} (2021) 128,
  [\href{http://arxiv.org/abs/1905.01311}{{\tt arXiv:1905.01311}}].

\bibitem{Manenti:2019kbl}
A.~Manenti, A.~Stergiou, and A.~Vichi, {\it {Implications of ANEC for SCFTs in
  four dimensions}},  {\em JHEP} {\bf 01} (2020) 093,
  [\href{http://arxiv.org/abs/1905.09293}{{\tt arXiv:1905.09293}}].

\bibitem{Belin:2020lsr}
A.~Belin, D.~M. Hofman, G.~Mathys, and M.~T. Walters, {\it {On the stress
  tensor light-ray operator algebra}},  {\em JHEP} {\bf 05} (2021) 033,
  [\href{http://arxiv.org/abs/2011.13862}{{\tt arXiv:2011.13862}}].

\bibitem{Besken:2020snx}
M.~Be\c{s}ken, J.~De~Boer, and G.~Mathys, {\it {On local and integrated
  stress-tensor commutators}},  {\em JHEP} {\bf 21} (2020) 148,
  [\href{http://arxiv.org/abs/2012.15724}{{\tt arXiv:2012.15724}}].

\bibitem{Korchemsky:2021okt}
G.~P. Korchemsky, E.~Sokatchev, and A.~Zhiboedov, {\it {Generalizing event
  shapes: in search of lost collider time}},  {\em JHEP} {\bf 08} (2022) 188,
  [\href{http://arxiv.org/abs/2106.14899}{{\tt arXiv:2106.14899}}].

\bibitem{Korchemsky:2021htm}
G.~P. Korchemsky and A.~Zhiboedov, {\it {On the light-ray algebra in conformal
  field theories}},  {\em JHEP} {\bf 02} (2022) 140,
  [\href{http://arxiv.org/abs/2109.13269}{{\tt arXiv:2109.13269}}].

\bibitem{Caron-Huot:2022eqs}
S.~Caron-Huot, M.~Kologlu, P.~Kravchuk, D.~Meltzer, and D.~Simmons-Duffin, {\it
  {Detectors in weakly-coupled field theories}},
  \href{http://arxiv.org/abs/2209.00008}{{\tt arXiv:2209.00008}}.

\bibitem{Kulaxizi:2010jt}
M.~Kulaxizi and A.~Parnachev, {\it {Energy Flux Positivity and Unitarity in
  CFTs}},  {\em Phys. Rev. Lett.} {\bf 106} (2011) 011601,
  [\href{http://arxiv.org/abs/1007.0553}{{\tt arXiv:1007.0553}}].

\bibitem{Karlsson:2022osn}
R.~Karlsson, A.~Parnachev, V.~Prilepina, and S.~Valach, {\it {Thermal stress
  tensor correlators, OPE and holography}},  {\em JHEP} {\bf 09} (2022) 234,
  [\href{http://arxiv.org/abs/2206.05544}{{\tt arXiv:2206.05544}}].

\bibitem{Maldacena:1997re}
J.~M. Maldacena, {\it {The Large N limit of superconformal field theories and
  supergravity}},  {\em Adv. Theor. Math. Phys.} {\bf 2} (1998) 231--252,
  [\href{http://arxiv.org/abs/hep-th/9711200}{{\tt hep-th/9711200}}].

\bibitem{Gubser:1998bc}
S.~S. Gubser, I.~R. Klebanov, and A.~M. Polyakov, {\it {Gauge theory
  correlators from noncritical string theory}},  {\em Phys. Lett. B} {\bf 428}
  (1998) 105--114, [\href{http://arxiv.org/abs/hep-th/9802109}{{\tt
  hep-th/9802109}}].

\bibitem{Witten:1998qj}
E.~Witten, {\it {Anti-de Sitter space and holography}},  {\em Adv. Theor. Math.
  Phys.} {\bf 2} (1998) 253--291,
  [\href{http://arxiv.org/abs/hep-th/9802150}{{\tt hep-th/9802150}}].

\bibitem{Brigante:2007nu}
M.~Brigante, H.~Liu, R.~C. Myers, S.~Shenker, and S.~Yaida, {\it {Viscosity
  Bound Violation in Higher Derivative Gravity}},  {\em Phys. Rev. D} {\bf 77}
  (2008) 126006, [\href{http://arxiv.org/abs/0712.0805}{{\tt
  arXiv:0712.0805}}].

\bibitem{Brigante:2008gz}
M.~Brigante, H.~Liu, R.~C. Myers, S.~Shenker, and S.~Yaida, {\it {The Viscosity
  Bound and Causality Violation}},  {\em Phys. Rev. Lett.} {\bf 100} (2008)
  191601, [\href{http://arxiv.org/abs/0802.3318}{{\tt arXiv:0802.3318}}].

\bibitem{deBoer:2009pn}
J.~de~Boer, M.~Kulaxizi, and A.~Parnachev, {\it {AdS(7)/CFT(6), Gauss-Bonnet
  Gravity, and Viscosity Bound}},  {\em JHEP} {\bf 03} (2010) 087,
  [\href{http://arxiv.org/abs/0910.5347}{{\tt arXiv:0910.5347}}].

\bibitem{Camanho:2009vw}
X.~O. Camanho and J.~D. Edelstein, {\it {Causality constraints in AdS/CFT from
  conformal collider physics and Gauss-Bonnet gravity}},  {\em JHEP} {\bf 04}
  (2010) 007, [\href{http://arxiv.org/abs/0911.3160}{{\tt arXiv:0911.3160}}].

\bibitem{Buchel:2009sk}
A.~Buchel, J.~Escobedo, R.~C. Myers, M.~F. Paulos, A.~Sinha, and M.~Smolkin,
  {\it {Holographic GB gravity in arbitrary dimensions}},  {\em JHEP} {\bf 03}
  (2010) 111, [\href{http://arxiv.org/abs/0911.4257}{{\tt arXiv:0911.4257}}].

\bibitem{Buchel:2009tt}
A.~Buchel and R.~C. Myers, {\it {Causality of Holographic Hydrodynamics}},
  {\em JHEP} {\bf 08} (2009) 016, [\href{http://arxiv.org/abs/0906.2922}{{\tt
  arXiv:0906.2922}}].

\bibitem{Buchel:2010wf}
A.~Buchel and S.~Cremonini, {\it {Viscosity Bound and Causality in Superfluid
  Plasma}},  {\em JHEP} {\bf 10} (2010) 026,
  [\href{http://arxiv.org/abs/1007.2963}{{\tt arXiv:1007.2963}}].

\bibitem{Cai:2010cv}
R.-G. Cai, Z.-Y. Nie, and H.-Q. Zhang, {\it {Holographic p-wave superconductors
  from Gauss-Bonnet gravity}},  {\em Phys. Rev. D} {\bf 82} (2010) 066007,
  [\href{http://arxiv.org/abs/1007.3321}{{\tt arXiv:1007.3321}}].

\bibitem{Bu:2015bwa}
Y.~Bu, M.~Lublinsky, and A.~Sharon, {\it {Hydrodynamics dual to
  Einstein-Gauss-Bonnet gravity: all-order gradient resummation}},  {\em JHEP}
  {\bf 06} (2015) 162, [\href{http://arxiv.org/abs/1504.01370}{{\tt
  arXiv:1504.01370}}].

\bibitem{Grozdanov:2016vgg}
S.~Grozdanov, N.~Kaplis, and A.~O. Starinets, {\it {From strong to weak
  coupling in holographic models of thermalization}},  {\em JHEP} {\bf 07}
  (2016) 151, [\href{http://arxiv.org/abs/1605.02173}{{\tt arXiv:1605.02173}}].

\bibitem{Andrade:2016yzc}
T.~Andrade, E.~Caceres, and C.~Keeler, {\it {Boundary causality versus
  hyperbolicity for spherical black holes in Gauss\textendash{}Bonnet
  gravity}},  {\em Class. Quant. Grav.} {\bf 34} (2017), no.~13 135003,
  [\href{http://arxiv.org/abs/1610.06078}{{\tt arXiv:1610.06078}}].

\bibitem{Grozdanov:2016zjj}
S.~Grozdanov and W.~van~der Schee, {\it {Coupling Constant Corrections in a
  Holographic Model of Heavy Ion Collisions}},  {\em Phys. Rev. Lett.} {\bf
  119} (2017), no.~1 011601, [\href{http://arxiv.org/abs/1610.08976}{{\tt
  arXiv:1610.08976}}].

\bibitem{Andrade:2016rln}
T.~Andrade, J.~Casalderrey-Solana, and A.~Ficnar, {\it {Holographic
  Isotropisation in Gauss-Bonnet Gravity}},  {\em JHEP} {\bf 02} (2017) 016,
  [\href{http://arxiv.org/abs/1610.08987}{{\tt arXiv:1610.08987}}].

\bibitem{Grozdanov:2016fkt}
S.~Grozdanov and A.~O. Starinets, {\it {Second-order transport, quasinormal
  modes and zero-viscosity limit in the Gauss-Bonnet holographic fluid}},  {\em
  JHEP} {\bf 03} (2017) 166, [\href{http://arxiv.org/abs/1611.07053}{{\tt
  arXiv:1611.07053}}].

\bibitem{Chen:2018nbh}
B.~Chen, P.-C. Li, Y.~Tian, and C.-Y. Zhang, {\it {Holographic Turbulence in
  Einstein-Gauss-Bonnet Gravity at Large $D$}},  {\em JHEP} {\bf 01} (2019)
  156, [\href{http://arxiv.org/abs/1804.05182}{{\tt arXiv:1804.05182}}].

\bibitem{An:2018dbz}
Y.-S. An, R.-G. Cai, and Y.~Peng, {\it {Time Dependence of Holographic
  Complexity in Gauss-Bonnet Gravity}},  {\em Phys. Rev. D} {\bf 98} (2018),
  no.~10 106013, [\href{http://arxiv.org/abs/1805.07775}{{\tt
  arXiv:1805.07775}}].

\bibitem{Grozdanov:2021gzh}
S.~Grozdanov, A.~O. Starinets, and P.~Tadi\'c, {\it {Hydrodynamic dispersion
  relations at finite coupling}},  {\em JHEP} {\bf 06} (2021) 180,
  [\href{http://arxiv.org/abs/2104.11035}{{\tt arXiv:2104.11035}}].

\bibitem{Camanho:2014apa}
X.~O. Camanho, J.~D. Edelstein, J.~Maldacena, and A.~Zhiboedov, {\it {Causality
  Constraints on Corrections to the Graviton Three-Point Coupling}},  {\em
  JHEP} {\bf 02} (2016) 020, [\href{http://arxiv.org/abs/1407.5597}{{\tt
  arXiv:1407.5597}}].

\bibitem{Kulaxizi:2017ixa}
M.~Kulaxizi, A.~Parnachev, and A.~Zhiboedov, {\it {Bulk Phase Shift, CFT Regge
  Limit and Einstein Gravity}},  {\em JHEP} {\bf 06} (2018) 121,
  [\href{http://arxiv.org/abs/1705.02934}{{\tt arXiv:1705.02934}}].

\bibitem{Li:2017lmh}
D.~Li, D.~Meltzer, and D.~Poland, {\it {Conformal Bootstrap in the Regge
  Limit}},  {\em JHEP} {\bf 12} (2017) 013,
  [\href{http://arxiv.org/abs/1705.03453}{{\tt arXiv:1705.03453}}].

\bibitem{Costa:2017twz}
M.~S. Costa, T.~Hansen, and J.~Penedones, {\it {Bounds for OPE coefficients on
  the Regge trajectory}},  {\em JHEP} {\bf 10} (2017) 197,
  [\href{http://arxiv.org/abs/1707.07689}{{\tt arXiv:1707.07689}}].

\bibitem{Bonifacio:2017nnt}
J.~Bonifacio, K.~Hinterbichler, A.~Joyce, and R.~A. Rosen, {\it {Massive and
  Massless Spin-2 Scattering and Asymptotic Superluminality}},  {\em JHEP} {\bf
  06} (2018) 075, [\href{http://arxiv.org/abs/1712.10020}{{\tt
  arXiv:1712.10020}}].

\bibitem{Afkhami-Jeddi:2018own}
N.~Afkhami-Jeddi, S.~Kundu, and A.~Tajdini, {\it {A Conformal Collider for
  Holographic CFTs}},  {\em JHEP} {\bf 10} (2018) 156,
  [\href{http://arxiv.org/abs/1805.07393}{{\tt arXiv:1805.07393}}].

\bibitem{Kologlu:2019bco}
M.~Kologlu, P.~Kravchuk, D.~Simmons-Duffin, and A.~Zhiboedov, {\it {Shocks,
  Superconvergence, and a Stringy Equivalence Principle}},  {\em JHEP} {\bf 11}
  (2020) 096, [\href{http://arxiv.org/abs/1904.05905}{{\tt arXiv:1904.05905}}].

\bibitem{Caron-Huot:2021enk}
S.~Caron-Huot, D.~Mazac, L.~Rastelli, and D.~Simmons-Duffin, {\it {AdS bulk
  locality from sharp CFT bounds}},  {\em JHEP} {\bf 11} (2021) 164,
  [\href{http://arxiv.org/abs/2106.10274}{{\tt arXiv:2106.10274}}].

\bibitem{Fitzpatrick:2019zqz}
A.~L. Fitzpatrick and K.-W. Huang, {\it {Universal Lowest-Twist in CFTs from
  Holography}},  {\em JHEP} {\bf 08} (2019) 138,
  [\href{http://arxiv.org/abs/1903.05306}{{\tt arXiv:1903.05306}}].

\bibitem{Zhiboedov:2013opa}
A.~Zhiboedov, {\it {On Conformal Field Theories With Extremal a/c Values}},
  {\em JHEP} {\bf 04} (2014) 038, [\href{http://arxiv.org/abs/1304.6075}{{\tt
  arXiv:1304.6075}}].

\bibitem{Boulware:1985wk}
D.~G. Boulware and S.~Deser, {\it {String Generated Gravity Models}},  {\em
  Phys. Rev. Lett.} {\bf 55} (1985) 2656.

\bibitem{Cai:2001dz}
R.-G. Cai, {\it {Gauss-Bonnet black holes in AdS spaces}},  {\em Phys. Rev. D}
  {\bf 65} (2002) 084014, [\href{http://arxiv.org/abs/hep-th/0109133}{{\tt
  hep-th/0109133}}].

\bibitem{Karlsson:2020ghx}
R.~Karlsson, M.~Kulaxizi, A.~Parnachev, and P.~Tadi\'c, {\it {Stress tensor
  sector of conformal correlators operators in the Regge limit}},  {\em JHEP}
  {\bf 07} (2020) 019, [\href{http://arxiv.org/abs/2002.12254}{{\tt
  arXiv:2002.12254}}].

\bibitem{Liu:1998bu}
H.~Liu and A.~A. Tseytlin, {\it {D = 4 superYang-Mills, D = 5 gauged
  supergravity, and D = 4 conformal supergravity}},  {\em Nucl. Phys. B} {\bf
  533} (1998) 88--108, [\href{http://arxiv.org/abs/hep-th/9804083}{{\tt
  hep-th/9804083}}].

\bibitem{El-Showk:2011yvt}
S.~El-Showk and K.~Papadodimas, {\it {Emergent Spacetime and Holographic
  CFTs}},  {\em JHEP} {\bf 10} (2012) 106,
  [\href{http://arxiv.org/abs/1101.4163}{{\tt arXiv:1101.4163}}].

\bibitem{Iliesiu:2018fao}
L.~Iliesiu, M.~Kolo\u{g}lu, R.~Mahajan, E.~Perlmutter, and D.~Simmons-Duffin,
  {\it {The Conformal Bootstrap at Finite Temperature}},  {\em JHEP} {\bf 10}
  (2018) 070, [\href{http://arxiv.org/abs/1802.10266}{{\tt arXiv:1802.10266}}].

\bibitem{Osborn:1993cr}
H.~Osborn and A.~C. Petkou, {\it {Implications of conformal invariance in field
  theories for general dimensions}},  {\em Annals Phys.} {\bf 231} (1994)
  311--362, [\href{http://arxiv.org/abs/hep-th/9307010}{{\tt hep-th/9307010}}].

\bibitem{Costa:2011mg}
M.~S. Costa, J.~Penedones, D.~Poland, and S.~Rychkov, {\it {Spinning Conformal
  Correlators}},  {\em JHEP} {\bf 11} (2011) 071,
  [\href{http://arxiv.org/abs/1107.3554}{{\tt arXiv:1107.3554}}].

\bibitem{Costa:2011dw}
M.~S. Costa, J.~Penedones, D.~Poland, and S.~Rychkov, {\it {Spinning Conformal
  Blocks}},  {\em JHEP} {\bf 11} (2011) 154,
  [\href{http://arxiv.org/abs/1109.6321}{{\tt arXiv:1109.6321}}].

\bibitem{Meyer:2011gj}
H.~B. Meyer, {\it {Transport Properties of the Quark-Gluon Plasma: A Lattice
  QCD Perspective}},  {\em Eur. Phys. J. A} {\bf 47} (2011) 86,
  [\href{http://arxiv.org/abs/1104.3708}{{\tt arXiv:1104.3708}}].

\bibitem{Karlsson:2019qfi}
R.~Karlsson, M.~Kulaxizi, A.~Parnachev, and P.~Tadi\'c, {\it {Black Holes and
  Conformal Regge Bootstrap}},  {\em JHEP} {\bf 10} (2019) 046,
  [\href{http://arxiv.org/abs/1904.00060}{{\tt arXiv:1904.00060}}].

\bibitem{Li:2019tpf}
Y.-Z. Li, Z.-F. Mai, and H.~L\"u, {\it {Holographic OPE Coefficients from AdS
  Black Holes with Matters}},  {\em JHEP} {\bf 09} (2019) 001,
  [\href{http://arxiv.org/abs/1905.09302}{{\tt arXiv:1905.09302}}].

\bibitem{Huang:2019fog}
K.-W. Huang, {\it {Stress-tensor commutators in conformal field theories near
  the lightcone}},  {\em Phys. Rev. D} {\bf 100} (2019), no.~6 061701,
  [\href{http://arxiv.org/abs/1907.00599}{{\tt arXiv:1907.00599}}].

\bibitem{Kulaxizi:2019tkd}
M.~Kulaxizi, G.~S. Ng, and A.~Parnachev, {\it {Subleading Eikonal, AdS/CFT and
  Double Stress Tensors}},  {\em JHEP} {\bf 10} (2019) 107,
  [\href{http://arxiv.org/abs/1907.00867}{{\tt arXiv:1907.00867}}].

\bibitem{Fitzpatrick:2019efk}
A.~L. Fitzpatrick, K.-W. Huang, and D.~Li, {\it {Probing universalities in d
  \ensuremath{>} 2 CFTs: from black holes to shockwaves}},  {\em JHEP} {\bf 11}
  (2019) 139, [\href{http://arxiv.org/abs/1907.10810}{{\tt arXiv:1907.10810}}].

\bibitem{Karlsson:2019dbd}
R.~Karlsson, M.~Kulaxizi, A.~Parnachev, and P.~Tadi\'c, {\it {Leading
  Multi-Stress Tensors and Conformal Bootstrap}},  {\em JHEP} {\bf 01} (2020)
  076, [\href{http://arxiv.org/abs/1909.05775}{{\tt arXiv:1909.05775}}].

\bibitem{Haehl:2019eae}
F.~M. Haehl, W.~Reeves, and M.~Rozali, {\it {Reparametrization modes, shadow
  operators, and quantum chaos in higher-dimensional CFTs}},  {\em JHEP} {\bf
  11} (2019) 102, [\href{http://arxiv.org/abs/1909.05847}{{\tt
  arXiv:1909.05847}}].

\bibitem{Li:2019zba}
Y.-Z. Li, {\it {Heavy-light Bootstrap from Lorentzian Inversion Formula}},
  {\em JHEP} {\bf 07} (2020) 046, [\href{http://arxiv.org/abs/1910.06357}{{\tt
  arXiv:1910.06357}}].

\bibitem{Karlsson:2019txu}
R.~Karlsson, {\it {Multi-stress tensors and next-to-leading singularities in
  the Regge limit}},  {\em JHEP} {\bf 08} (2020) 037,
  [\href{http://arxiv.org/abs/1912.01577}{{\tt arXiv:1912.01577}}].

\bibitem{Huang:2020ycs}
K.-W. Huang, {\it {Lightcone Commutator and Stress-Tensor Exchange in $d>2$
  CFTs}},  {\em Phys. Rev. D} {\bf 102} (2020), no.~2 021701,
  [\href{http://arxiv.org/abs/2002.00110}{{\tt arXiv:2002.00110}}].

\bibitem{Li:2020dqm}
Y.-Z. Li and H.-Y. Zhang, {\it {More on heavy-light bootstrap up to
  double-stress-tensor}},  {\em JHEP} {\bf 10} (2020) 055,
  [\href{http://arxiv.org/abs/2004.04758}{{\tt arXiv:2004.04758}}].

\bibitem{Parnachev:2020fna}
A.~Parnachev, {\it {Near Lightcone Thermal Conformal Correlators and
  Holography}},  {\em J. Phys. A} {\bf 54} (2021), no.~15 155401,
  [\href{http://arxiv.org/abs/2005.06877}{{\tt arXiv:2005.06877}}].

\bibitem{Fitzpatrick:2020yjb}
A.~L. Fitzpatrick, K.-W. Huang, D.~Meltzer, E.~Perlmutter, and
  D.~Simmons-Duffin, {\it {Model-dependence of minimal-twist OPEs in d
  \ensuremath{>} 2 holographic CFTs}},  {\em JHEP} {\bf 11} (2020) 060,
  [\href{http://arxiv.org/abs/2007.07382}{{\tt arXiv:2007.07382}}].

\bibitem{Berenstein:2020vlp}
D.~Berenstein, Z.~Li, and J.~Simon, {\it {ISCOs in AdS/CFT}},  {\em Class.
  Quant. Grav.} {\bf 38} (2021), no.~4 045009,
  [\href{http://arxiv.org/abs/2009.04500}{{\tt arXiv:2009.04500}}].

\bibitem{Parnachev:2020zbr}
A.~Parnachev and K.~Sen, {\it {Notes on AdS-Schwarzschild eikonal phase}},
  {\em JHEP} {\bf 03} (2021) 289, [\href{http://arxiv.org/abs/2011.06920}{{\tt
  arXiv:2011.06920}}].

\bibitem{Karlsson:2021duj}
R.~Karlsson, A.~Parnachev, and P.~Tadi\'c, {\it {Thermalization in large-N
  CFTs}},  {\em JHEP} {\bf 09} (2021) 205,
  [\href{http://arxiv.org/abs/2102.04953}{{\tt arXiv:2102.04953}}].

\bibitem{Rodriguez-Gomez:2021pfh}
D.~Rodriguez-Gomez and J.~G. Russo, {\it {Correlation functions in finite
  temperature CFT and black hole singularities}},  {\em JHEP} {\bf 06} (2021)
  048, [\href{http://arxiv.org/abs/2102.11891}{{\tt arXiv:2102.11891}}].

\bibitem{Huang:2021hye}
K.-W. Huang, {\it {$d>2$ stress-tensor operator product expansion near a
  line}},  {\em Phys. Rev. D} {\bf 103} (2021), no.~12 121702,
  [\href{http://arxiv.org/abs/2103.09930}{{\tt arXiv:2103.09930}}].

\bibitem{Rodriguez-Gomez:2021mkk}
D.~Rodriguez-Gomez and J.~G. Russo, {\it {Thermal correlation functions in CFT
  and factorization}},  {\em JHEP} {\bf 11} (2021) 049,
  [\href{http://arxiv.org/abs/2105.13909}{{\tt arXiv:2105.13909}}].

\bibitem{Krishna:2021fus}
H.~Krishna and D.~Rodriguez-Gomez, {\it {Holographic thermal correlators
  revisited}},  {\em JHEP} {\bf 11} (2021) 139,
  [\href{http://arxiv.org/abs/2108.00277}{{\tt arXiv:2108.00277}}].

\bibitem{Bianchi:2021yqs}
M.~Bianchi and G.~Di~Russo, {\it {Turning black holes and D-branes inside out
  of their photon spheres}},  {\em Phys. Rev. D} {\bf 105} (2022), no.~12
  126007, [\href{http://arxiv.org/abs/2110.09579}{{\tt arXiv:2110.09579}}].

\bibitem{Karlsson:2021mgg}
R.~Karlsson, M.~Kulaxizi, G.~S. Ng, A.~Parnachev, and P.~Tadi\'c, {\it {CFT
  correlators, $ \mathcal{W} $-algebras and generalized Catalan numbers}},
  {\em JHEP} {\bf 06} (2022) 162, [\href{http://arxiv.org/abs/2111.07924}{{\tt
  arXiv:2111.07924}}].

\bibitem{Huang:2022vcs}
K.-W. Huang, {\it {Approximate symmetries in d = 4 CFTs with an Einstein
  gravity dual}},  {\em JHEP} {\bf 09} (2022) 053,
  [\href{http://arxiv.org/abs/2202.09998}{{\tt arXiv:2202.09998}}].

\bibitem{Dodelson:2022eiz}
M.~Dodelson and A.~Zhiboedov, {\it {Gravitational orbits, double-twist mirage,
  and many-body scars}},  \href{http://arxiv.org/abs/2204.09749}{{\tt
  arXiv:2204.09749}}.

\bibitem{Dodelson:2022yvn}
M.~Dodelson, A.~Grassi, C.~Iossa, D.~Panea~Lichtig, and A.~Zhiboedov, {\it
  {Holographic thermal correlators from supersymmetric instantons}},
  \href{http://arxiv.org/abs/2206.07720}{{\tt arXiv:2206.07720}}.

\bibitem{Parisini:2022wkb}
E.~Parisini, K.~Skenderis, and B.~Withers, {\it {An embedding formalism for
  CFTs in general states on curved backgrounds}},
  \href{http://arxiv.org/abs/2209.09250}{{\tt arXiv:2209.09250}}.

\bibitem{Z1985}
A.~B. Zamolodchikov, {\it {Infinite additional symmetries in two-dimensional
  conformal quantum field theory}},  {\em Theoretical and Mathematical Physics}
  {\bf 65} (1985).

\bibitem{lovelock1971}
D.~Lovelock, {\it {The Einstein tensor and its generalizations}},  {\em J.
  Math. Phys.} {\bf 12} (1971) 498--501.

\end{thebibliography}\endgroup

\end{document}